\documentclass[showpacs,preprintnumbers,amsmath,amssymb]{revtex4}
\usepackage{graphicx}
\usepackage{dcolumn}
\usepackage{bm}
\usepackage{epsf}
\newcommand{\p}{\partial}
\newcommand{\omu}{\overline{\mu}}

\newcommand{\occ}{\overline{c}}
\newcommand{\al}{\alpha}

\newcommand{\lms}{\Lambda_{\overline{\mathrm{MS}}}}

\newcommand{\MSbar}{\overline{\mathrm{MS}}}

\newcommand{\Evac}{E_{\textrm{vac}}}
\newcommand{\oo}{\mathcal{O}}
\newcommand{\wsigma}{\widetilde{\sigma}}
\newcommand{\os}{\overline{s}}
\newcommand{\mw}{\mathcal{W}}
\newcommand{\mf}{\mathcal{F}}
\newcommand{\s}{\sigma}
\newcommand{\wj}{\widetilde{J}}
\begin{document}
\preprint{LTH-623}
\title{An analytic study of the off-diagonal mass generation for Yang-Mills theories in the maximal Abelian gauge}
\author{D. Dudal$^a$}
 \altaffiliation{Research Assistant of the Fund For Scientific Research-Flanders (Belgium)}
\email{david.dudal@ugent.be}
\author{J.A. Gracey$^b$}
\email{jag@amtp.liv.ac.uk}
\author{V.E.R. Lemes$^c$}
  \email{vitor@dft.if.uerj.br}
\author{M.S. Sarandy$^d$}
\email{msarandy@chem.utoronto.ca}
\author{R.F. Sobreiro$^c$}
 \email{sobreiro@uerj.br}
\author{S.P. Sorella$^c$}
\email{sorella@uerj.br}
\author{H. Verschelde$^a$}
 \email{henri.verschelde@ugent.be}
 \affiliation{\vskip 0.1cm $^a$ Ghent University
\\ Department of Mathematical
Physics and Astronomy \\ Krijgslaan 281-S9 \\ B-9000 Gent,
Belgium\\\\ \vskip 0.1cm $^b$ Theoretical Physics Division \\
Department of Mathematical Sciences \\
University of Liverpool \\
P.O. Box 147, Liverpool, L69 3BX, United Kingdom\\\\
\vskip 0.1cm $^c$ UERJ - Universidade do Estado do Rio de
Janeiro\\Rua S\~{a}o Francisco Xavier 524, 20550-013
Maracan\~{a}\\Rio de Janeiro, Brazil\\\\ \vskip 0.1cm $^d$
Chemical Physics Theory Group, Department of Chemistry, University of Toronto, 80 St. George Street, Toronto, Ontario, M5S 3H6, Canada}%
\begin{abstract}
We investigate a dynamical mass generation mechanism for the
off-diagonal gluons and ghosts in $SU(N)$ Yang-Mills theories,
quantized in the maximal Abelian gauge. Such a mass can be seen as
evidence for the Abelian dominance in that gauge. It originates from
the condensation of a mixed gluon-ghost operator of mass dimension
two, which lowers the vacuum energy. We construct an effective
potential for this operator by a combined use of the local composite
operators technique with the algebraic renormalization and we
discuss the gauge parameter independence of the results. We also
show that it is possible to connect the vacuum energy, due to the
mass dimension two condensate discussed here, with the non-trivial
vacuum energy originating from the condensate $\left\langle A_\mu^2
\right\rangle$, which has attracted much attention in the Landau
gauge.
\end{abstract}
 \maketitle
\section{\label{sec1}Introduction.}
A widely accepted mechanism to explain color confinement in $SU(N)$
Yang-Mills theories is based on the dual superconductivity picture
\cite{scon,'tHooft:1981ht},  according to which the low energy
regime of QCD should be described by an effective Abelian theory in
the presence of magnetic monopoles. These monopoles should condense,
giving rise to a string formation $\grave{\textrm{a}}$ la
Abrikosov-Nielsen-Olesen. As a result, chromoelectric charges are
confined. This mechanism has received many confirmations from the
lattice community in the so-called Abelian gauges, which are useful
in order to  isolate the effective relevant degrees of freedom at
low energy.

According to the concept of Abelian dominance, the low energy
region of QCD can be expressed solely in terms of Abelian degrees
of freedom \cite{Ezawa:bf}. Lattice confirmations of the Abelian
dominance can be found in \cite{Suzuki:1989gp,Hioki:1991ai}. A
particularly interesting Abelian gauge is the maximal Abelian
gauge (MAG), introduced in
\cite{'tHooft:1981ht,Kronfeld:1987vd,Kronfeld:1987ri}. Roughly
speaking, the MAG  is obtained by minimizing the square of the
norm of the fields corresponding to off-diagonal gluons, i.e. the
gluons associated with the $N(N-1)$ off-diagonal generators of
$SU(N)$. Doing so, there is a residual $U(1)^{N-1}$ Abelian gauge
freedom corresponding to the Cartan subgroup of $SU(N)$. The
renormalizability in the continuum of this gauge was proven in
\cite{Min:bx,Fazio:2001rm}, at the cost of introducing a quartic
ghost interaction.

To our knowledge, there is no analytic proof  of the Abelian
dominance. Nevertheless, an argument that can be interpreted  as
evidence of it, is the fact that the off-diagonal gluons would
attain a dynamical mass. At energies below the scale set by this
mass, the off-diagonal gluons should decouple, and in this way one
should end up with an Abelian theory at low energies. A lattice
study of such an off-diagonal gluon mass reported a value of
approximately $1.2$GeV \cite{Amemiya:1998jz}. More recently, the
off-diagonal gluon propagator was investigated numerically in
\cite{Bornyakov:2003ee}, reporting a similar result.

There have been  several efforts to give an analytic  description of
the mechanism  responsible for the dynamical generation of the
off-diagonal gluon mass. In \cite{Schaden:1999ew,Kondo:2000ey}, a
certain ghost condensate was used to construct an effective,
off-diagonal mass. However, in \cite{Dudal:2002xe} it was shown that
the obtained mass was a tachyonic one, a fact confirmed later in
\cite{Sawayanagi:2003dc}. Another condensation, namely that of the
mixed gluon-ghost operator $(\frac{1}{2}A_\mu^a A^{\mu a}+\al\occ^a
c^a)$ \footnote{The index $a$ runs only over the $N(N-1)$
off-diagonal generators.}, that could be responsible for the
off-diagonal mass, was proposed in \cite{Kondo:2001nq}.
 That this operator should condense can be expected on
the basis of a close analogy existing between the MAG and the
renormalizable nonlinear Curci-Ferrari gauge
\cite{Curci:bt,Curci:1976ar}. In fact, it turns out that the mixed
gluon-ghost operator can be introduced also in the Curci-Ferrari
gauge. A detailed analysis of its condensation and of the ensuing
dynamical mass generation can be found in
\cite{Dudal:2003pe,Dudal:2003gu}.

The aim of this paper is to investigate explicitly if the mass
dimension two operator $(\frac{1}{2}A_\mu^a A^{\mu a}+\al\occ^a
c^a)$ condenses, so that a dynamical off-diagonal mass is generated
in the MAG. The pathway we intend to follow is based on previous
research in this direction in other gauges. In
\cite{Verschelde:2001ia}, the local composite operator (LCO)
technique was used to construct a renormalizable effective potential
for the operator $A_\mu^A A^{\mu A}$ in the Landau gauge. As a
consequence of $\left\langle A_\mu^A A^{\mu A}\right\rangle\neq0$,
the gauge bosons acquired a mass \cite{Verschelde:2001ia}. The fact
that gluons in the Landau gauge become massive  has received
confirmations from lattice simulations, see for example
\cite{Langfeld:2001cz}. Recently, the dynamical mass generation in
the Landau gauge has been investigated within the Schwinger-Dyson
formalism in \cite{Aguilar:2004kt,Aguilar:2004sw}. The condensate
$\left\langle A_\mu^A A^{\mu A}\right\rangle$ has attracted
attention from theoretical \cite{Gubarev:2000eu,Gubarev:2000nz} as
well as from the lattice side
\cite{Boucaud:2000nd,Boucaud:2001st,Boucaud:2002nc}. It was shown by
means of the algebraic renormalization technique \cite{book} that
the LCO formalism  for the condensate $\left\langle A_\mu^A A^{\mu
A}\right\rangle$ is renormalizable to all orders of perturbation
theory \cite{Dudal:2002pq}.  The same formalism was successfully
employed to study the condensation of $(\frac{1}{2}A_\mu^A A^{\mu
A}+\al\occ^A c^A)$ in the Curci-Ferrari gauge
\cite{Dudal:2003pe,Dudal:2003gu}. We would like to note that the
Landau gauge corresponds to $\al=0$. Later on, the condensation of
$A_\mu^A A^{\mu A}$ was confirmed in the linear covariant gauges
\cite{Dudal:2003np,Dudal:2003by}, which also possess the Landau
gauge as a special case. It was proven formally that the vacuum
energy does not depend on the gauge parameter. However, in practice,
a problem occurred due to the mixing of  different orders of
perturbation theory, when solving the gap equation for the
condensate.  Nevertheless, we have been able to present a way to
overcome this problem \cite{Dudal:2003by}. As a result,  it turns
out that the non-trivial vacuum energy due to the condensate
$\left\langle A_\mu^A A^{\mu A}\right\rangle$ in the Landau gauge
coincides with the non-trivial vacuum energy due to the appropriate
mass dimension two condensate in the linear covariant gauges,
 $\left\langle A_\mu^A A^{\mu A}\right\rangle$, and
the Curci-Ferrari gauge, $\left\langle\frac{1}{2}A_\mu^A A^{\mu
A}+\al\occ^A c^A\right\rangle$, since these two classes of gauges
both have the Landau gauge,  $\alpha=0$, as a limiting case.

We would also like to underline that the concept of a gluon mass has
already been widely used in a more phenomenological context since
long ago, see e.g. \cite{Parisi:1980jy}. More recently, a gluon mass
of the order of a few hundred MeV has been proven to be very useful
in describing the radiative decay of heavy quarkonia systems
\cite{Field:2001iu} as well as to derive estimates of the glueball
spectrum \cite{Giacosa:2004ug}.

To make this paper self-contained, we will explain all necessary
steps in the case of the MAG, and refer to the previous papers for
more details where appropriate. In section II, we introduce the
MAG and discuss its renormalizability when the operator
$(\frac{1}{2}A_\mu^a A^{\mu a}+\al\occ^a c^a)$ is introduced in
the theory. We briefly review how the effective potential is
constructed by means of the LCO technique. In section III,  we
discuss the independence of the vacuum energy from the gauge
parameter of the MAG. We face the problem of the mixing of
different orders in perturbation theory, and we provide a solution
of it. In section IV, we construct a  generalized renormalizable
gauge that  interpolates between the MAG and the Landau gauge.
Moreover, we will also show that there exists a generalized
renormalizable mass dimension two operator that interpolates
between the mass dimension two operators of the MAG and of the
Landau gauge. This can be used to prove that the vacuum energy
obtained in the MAG is the same as  that of the Landau gauge. In
section V, we  present explicit results, obtained in the case of
$SU(2)$ and to the one loop approximation. We end with conclusions
in section VI.
\section{$SU(N)$ Yang-Mills theories in the MAG.}
Let $A_{\mu }$ be the Lie algebra valued connection for the gauge
group $SU(N)$, whose generators $T^{A}$, satisfying $\left[
T^{A},T^{B}\right]$~$=$~$f^{ABC}T^{C}$, are chosen to be
antihermitean and to obey the
orthonormality condition $\mathrm{Tr}\left( T^{A}T^{B}\right) =-T_F\delta ^{AB}$%
, with $A,B,C=1,\ldots,\left( N^{2}-1\right) $. In the case of
$SU(N)$, one has $T_F=\frac{1}{2}$. We decompose the gauge field
into its off-diagonal and diagonal parts, namely
\begin{equation}
A_{\mu }=A_{\mu }^{A}T^{A}=A_{\mu }^{a}T^{a}+A_{\mu }^{i}T^{\,i},
\label{conn}
\end{equation}
where  the indices $i$, $j$, $\ldots$ label the $N-1$ generators
of the Cartan subalgebra. The remaining $N(N-1)$ off-diagonal
generators will be labelled by the indices $a$, $b$, $\ldots$. For
further  use, we recall the Jacobi identity
\begin{equation}\label{jacobi}
    f^{ABC}f^{CDE}+f^{ADC}f^{CEB}+f^{AEC}f^{CBD}=0\;,
\end{equation}
from which it can be deduced that
\begin{eqnarray}\label{jacobi2}
f^{abi}f^{bjc}+f^{abj}f^{bci}&=&0\;,\nonumber\\
f^{abc}f^{bdi}+f^{abd}f^{bic}+f^{abi}f^{bcd}&=&0\;.
\end{eqnarray}
The field strength decomposes as
\begin{equation}
F_{\mu \nu }=F_{\mu \nu }^{A}T^{A}=F_{\mu \nu }^{a}T^{a}+F_{\mu
\nu }^{i}T^{\,i}\;,  \label{fs}
\end{equation}
with the off-diagonal and diagonal parts given respectively by
\begin{eqnarray}
F_{\mu \nu }^{a} &=&D_{\mu }^{ab}A_{\nu }^{b}-D_{\nu }^{ab}A_{\mu
}^{b}\;+g\,f^{abc}A_{\mu }^{b}A_{\nu }^{c}\;,  \label{fsc} \\
F_{\mu \nu }^{i} &=&\partial _{\mu }A_{\nu }^{i}-\partial _{\nu
}A_{\mu }^{i}+gf^{abi}A_{\mu }^{a}A_{\nu }^{b}\;,  \nonumber
\end{eqnarray}
where the covariant derivative $D_{\mu }^{ab}$ is defined with
respect to the diagonal components $A_{\mu }^{i}$
\begin{equation}
D_{\mu }^{ab}\equiv \partial _{\mu }\delta ^{ab}-gf^{abi}A_{\mu
}^{i}\,\,\,\,\,\,.  \label{cv}
\end{equation}
For the Yang-Mills action one obtains
\begin{equation}
S_{\mathrm{YM}}=-\frac{1}{4}\int d^{4}x\,\left( F_{\mu \nu
}^{a}F^{\mu \nu a}+F_{\mu \nu }^{i}F^{\mu \nu i}\right) \;.
\label{ym}
\end{equation}
The so called MAG gauge condition amounts to fixing the value of
the covariant derivative, $D_{\mu }^{ab}A^{\mu b}$, of the
off-diagonal components by requiring that the functional
\begin{equation}\label{MAGfunctional}
    \mathcal{R}[A]=(VT)^{-1}\int d^4x\left(A_\mu^a A^{\mu a}\right)\;,
\end{equation}
 attains a minimum with respect to the local gauge
transformations.  This corresponds to imposing
\begin{equation}\label{MAGdiff}
    D_\mu^{ab}A^{\mu b}=0\;.
\end{equation}
However, this condition being non-linear implies a quartic ghost
self-interaction term is required for renormalizability purposes.
The corresponding gauge fixing term turns out to be
\cite{Min:bx,Fazio:2001rm}
\begin{equation}
S_{\mathrm{MAG}}=s\,\int d^{4}x\,\left( \overline{c}^{a}\left(
D_{\mu
}^{ab}A^{b\mu }+\frac{\alpha }{2}b^{a}\right) -\frac{\alpha }{2}gf\,^{abi}%
\overline{c}^{a}\overline{c}^{b}c^{i}-\frac{\alpha }{4}gf\,^{abc}c^{a}%
\overline{c}^{b}\overline{c}^{c}\right) \;,  \label{smn}
\end{equation}
where $\alpha $ is the MAG gauge parameter and $s$ denotes the
nilpotent BRST operator, acting as
\begin{eqnarray}
sA_{\mu }^{a} &=&-\left( D_{\mu }^{ab}c^{b}+gf^{\,abc}A_{\mu
}^{b}c^{c}+gf^{\,abi}A_{\mu }^{b}c^{i}\right) ,\,\,\,\;sA_{\mu
}^{i}=-\left(
\partial _{\mu }c^{i}+gf\,^{iab}A_{\mu }^{a}c^{b}\right) \;,  \nonumber \\
sc^{a} &=&gf\,^{abi}c^{b}c^{i}+\frac{g}{2}f\,^{abc}c^{b}c^{c},\,\,\,\,\,\,\,%
\,\,\,\,\,\,\,\,\,\,\,\,\,\,\,\,\,\,\,\,\,\,\,\,\,\,\,\,\,\,\,\,\,\,\,sc^{i}=%
\frac{g}{2}\,f\,^{iab}c^{a}c^{b},  \nonumber \\
s\overline{c}^{a}
&=&b^{a}\;,\,\,\,\,\,\,\,\,\,\,\,\,\,\,\,\,\,\,\,\,\,\,\,\,\,\,\,\,\,\,\,\,%
\,\,\,\,\,\,\,\,\,\,\,\,\,\,\,\,\,\,\,\,\,\,\,\,\,\,\,\,\,\,\,\,\,\,\,\,\,\,%
\,\,\,\,\,\,\,\,\,\,\,\,\,\,\,\,\,\,\,\,\,\,s\overline{c}^{i}=b^{i}\;,  \nonumber \\
sb^{a}
&=&0\;,\,\,\,\,\,\,\,\,\,\,\,\,\,\,\,\,\,\,\,\,\,\,\,\,\,\,\,\,\,\,\,\,\,\,%
\,\,\,\,\,\,\,\,\,\,\,\,\,\,\,\,\,\,\,\,\,\,\,\,\,\,\,\,\,\,\,\,\,\,\,\,\,\,%
\,\,\,\,\,\,\,\,\,\,\,\,\,\,\,\,\,\,\,\,\,\,sb^{i}=0\;.
\label{brst}
\end{eqnarray}
Here $c^{a},c^{i}$ are the off-diagonal and the diagonal
components of the Faddeev-Popov ghost field, while
$\overline{c}^{a},b^{a}$ are the off-diagonal antighost and
Lagrange multiplier. We also observe that the BRST\
transformations $\left( \ref{brst}\right) $ have been obtained by
their standard form upon projection on the off-diagonal and
diagonal components of the fields.  We remark that the MAG
(\ref{smn}) can be written in the form
\begin{equation}
S_{\mathrm{MAG}}=s\os\int d^{4}x\,\left(\frac{1}{2}A_\mu^a A^{\mu
a}-\frac{\al}{2}c^a\occ^a\right) \;, \label{smn2}
\end{equation}
with $\os$  being the nilpotent anti-BRST transformation, acting
as
\begin{eqnarray}
\os A_{\mu }^{a} &=&-\left( D_{\mu }^{ab}\occ^{b}+gf^{\,abc}A_{\mu
}^{b}\occ^{c}+gf^{\,abi}A_{\mu }^{b}\occ^{i}\right)
,\,\,\,\,\,\,\,\;\os A_{\mu }^{i}=-\left(
\partial _{\mu }\occ^{i}+gf\,^{iab}A_{\mu }^{a}\occ^{b}\right) \;,  \nonumber \\
\os\occ^{a} &=&gf^{abi}\occ^{b}\occ^{i}+\frac{g}{2}f^{abc}\occ^{b}\occ^{c},\,\,\,\,\,\,\,%
\,\,\,\,\,\,\,\,\,\,\,\,\,\,\,\,\,\,\,\,\,\,\,\,\,\,\,\,\,\,\,\,\,\,\,\,\,\,\,\,\,\,\,\os\occ^{i}=%
\frac{g}{2}\,f\,^{iab}\occ^{a}\occ^{b},  \nonumber \\
\os c^{a}
&=&-b^{a}+gf^{abc}c^b\occ^c+gf^{abi}c^b\occ^i+gf^{abi}\occ^bc^i\;,\,\,\,\,\os c^{i}=-b^{i}+gf^{ibc}c^b\occ^c\;,  \nonumber \\
\os b^{a} &=&-gf^{abc}b^b\occ^c-gf^{abi}b^b\occ^i+gf^{abi}\occ^b
b^i\;\;\;\;\;\;\;\;\;\;\;\;\;\os b^i=-gf^{ibc}b^b\occ^c\;.
\label{brst2}
\end{eqnarray}
It can be checked that $s$ and $\os$ anticommute.

Expression $\left( \ref{smn}\right) $ is easily worked out and
yields
\begin{eqnarray}
S_{\mathrm{MAG}} &=&\int d^{4}x\left( b^{a}\left( D_{\mu }^{ab}A^{\mu b}+%
\frac{\alpha }{2}b^{a}\right) +\overline{c}^{a}D_{\mu }^{ab}D^{\mu bc}c^{c}+g%
\overline{c}^{a}f^{abi}\left( D_{\mu }^{bc}A^{\mu c}\right)
c^{i}+g\overline{c}^{a}D_{\mu }^{ab}\left( f^{bcd}A^{\mu c
}c^{d}\right) \right.\nonumber\\&-&\alpha
gf^{abi}b^{a}\overline{c}^{b}c^{i}-\left.g^{2}f^{abi}f^{cdi}\overline{c}^{a}c^{d}A_{\mu
}^{b}A^{\mu c}-\frac{\alpha }{2}gf^{abc}b^{a}\overline{c}%
^{b}c^{c}-\frac{\alpha }{4}g^{2}f^{abi}f^{cdi}\overline{c}^{a}\overline{c}%
^{b}c^{c}c^{d}\right.\nonumber\\&-&\left.\frac{\alpha }{4}g^{2}f^{abc}f^{adi}\overline{c}^{b}\overline{c}%
^{c}c^{d}c^{i}  -\frac{\alpha }{8}g^{2}f^{abc}f^{ade}%
\overline{c}^{b}\overline{c}^{c}c^{d}c^{e}\right)\;. \label{smne}
\end{eqnarray}
We note that $\al=0$ does in fact correspond to the ``real'' MAG
condition, given by eq.(\ref{MAGdiff}). However, one cannot set
$\al=0$ from the beginning since this would lead to a
nonrenormalizable gauge. Some of the terms proportional to $\al$
would reappear due to radiative corrections, even if $\al=0$. See,
for example, \cite{Kondo:1997pc}. For our purposes, this means
that we have to keep $\al$ general  throughout and leave to the
end the analysis of the limit $\al\rightarrow 0$, to recover
condition (\ref{MAGdiff}).

The MAG condition allows for a residual local $U(1)^{N-1}$
invariance with respect to the diagonal subgroup. In order to have
a complete quantization of the theory, one has to fix this Abelian
gauge freedom by means of a suitable further gauge condition on
the diagonal components $A_{\mu }^{i}$ of the gauge field.  A
common choice for the Abelian gauge fixing, also adopted in the
lattice papers \cite{Amemiya:1998jz,Bornyakov:2003ee}, is the
Landau gauge, given by
\begin{equation}
S_{\mathrm{diag}}=s\,\int d^{4}x\,\;\overline{c}^{i}\partial _{\mu
}A^{\mu i
}\;=\int d^{4}x\,\;\left( b^{i}\partial _{\mu }A^{\mu i}+\overline{c}%
^{i}\partial ^{\mu }\left( \partial _{\mu }c^{i}+gf\,^{iab}A_{\mu
}^{a}c^{b}\right) \right) \;,  \label{abgf}
\end{equation}
where $\overline{c}^{i},b^{i}$ are the diagonal antighost and
Lagrange multiplier.
\subsection{Ward identities for the MAG.}
In order to write down a suitable set of Ward identities, we first
introduce external fields $\Omega^{\mu i}$, $\Omega^{\mu a}$,
$L^{i}$, $L^{a}$ coupled to the BRST\ nonlinear variations of the
fields, namely
\begin{eqnarray}
S_{\mathrm{ext}} &=&\int d^{4}x\left( -\Omega ^{\mu a}\left(
D_{\mu }^{ab}c^{b}+gf^{abc}A_{\mu }^{b}c^{c}+gf^{abi}A_{\mu
}^{b}c^{i}\right)  -\Omega^{\mu i}\left( \partial _{\mu
}c^{i}+gf^{iab}A_{\mu }^{a}c^{b}\right)\right.\nonumber\\ &+&\left.L^{a}\left( gf^{abi}c^{b}c^{i}+%
\frac{g}{2}f^{abc}c^{b}c^{c}\right)
+L^{i}\frac{g}{2}\,f^{iab}c^{a}c^{b}\right) \;,   \label{sexr}
\end{eqnarray}
with
\begin{eqnarray}
s\Omega^{\mu a} &=&s\Omega^{\mu i}=0\;,  \label{ss} \\
sL^{a} &=&sL^{i}=0\;.  \nonumber
\end{eqnarray}
Moreover, in order to discuss the renormalizability of the
gluon-ghost operator
\begin{equation}
\mathcal{O}_{\mathrm{MAG}}= \frac{1}{2}A_{\mu}^{a}A^{\mu a}+\alpha \overline{c}%
^{a}c^{a} \;,  \label{ggop}
\end{equation}
we introduce it in the starting action by means of a BRST\ doublet
of external sources $\left( J,\lambda \right)$
\begin{equation}
s\lambda =J\;,\;\;\;\;sJ=0\;,  \label{jl}
\end{equation}
so that
\begin{eqnarray}
S_{\mathrm{LCO}} &=&s\int d^{4}x\;\left( \lambda \left(
\frac{1}{2}A_{\mu
}^{a}A^{\mu a}+\alpha \overline{c}^{a}c^{a}\right) +\zeta \frac{\lambda J}{2%
}\right) \;  \label{slco} \\
&=&\int d^{4}x\;\left( J\left( \frac{1}{2}A_\mu^a A^{\mu a}+\alpha
\overline{c}^{a}c^{a}\right) +\zeta \frac{J^{2}}{2}-\alpha \lambda
b^{a}c^{a}\right.   \nonumber \\
&+&\left. \lambda A^{\mu a}\left( D_{\mu
}^{ab}c^{b}+gf^{abi}A_{\mu }^{b}c^{i}\right) +\alpha \lambda
\overline{c}^{a}\left(
gf\,^{abi}c^{b}c^{i}+\frac{g}{2}f\,^{abc}c^{b}c^{c}\right) \right)
\;, \nonumber
\end{eqnarray}
where $\zeta $ is the LCO parameter accounting for the divergences
present in the vacuum correlator
$\left\langle\mathcal{O}_{\mathrm{MAG}}(x)\mathcal{O}_{\mathrm{MAG}}(y)
\right\rangle $, which are proportional to $J^{2}$. Therefore, the
complete action
\begin{equation}
\Sigma =S_{\mathrm{YM}}+S_{\mathrm{MAG}}+S_{\mathrm{diag}}+S_{\mathrm{ext}%
}+S_{\mathrm{LCO}}\;,  \label{ca}
\end{equation}
is BRST invariant
\begin{equation}
s\Sigma =0\;.  \label{inv}
\end{equation}
As noticed in \cite{Kondo:2001nq,Kondo:2001tm}, the gluon-ghost
mass operator defined in eq.(\ref{ggop}) is BRST invariant
on-shell.

In Table I, the dimension and ghost number of all the fields and
sources are listed.
\begin{table}[t]
  \centering
  \begin{tabular}{|c|c|c|c|c|c|c|c|c|}
\hline
    &  $A_\mu^{a,i}$ & $c^{a,i}$ & $\occ^{a,i}$ & $b^{a,i}$ & $\lambda$ & $J$ & $\Omega_\mu^{a,i}$ & $L^{a,i}$ \\
    \hline
    dimension & 1 & 0 & 2 & 2 & 2 & 2 & 3 & 4 \\
    ghost number & 0 & 1 & $-1$ & 0 & $-1$ & 0 & $-1$ & $-2$ \\ \hline
  \end{tabular}
  \caption{Dimension and ghost number.}\label{table1}
\end{table}
We are now ready to write down the Ward identities needed to
discuss the renormalizability of the model. It turns out that the complete action $%
\Sigma $ is constrained by
\begin{itemize}
\item  the Slavnov-Taylor identity
\begin{equation}
\mathcal{S}(\Sigma )=0\;,  \label{st}
\end{equation}
with
\begin{eqnarray}
\mathcal{S}(\Sigma ) &=&\int d^{4}x\left( \frac{\delta \Sigma
}{\delta \Omega ^{\mu a}}\frac{\delta \Sigma }{\delta A_{\mu
}^{a}}+\frac{\delta
\Sigma }{\delta \Omega ^{\mu i}}\frac{\delta \Sigma }{\delta A_{\mu }^{i}}+%
\frac{\delta \Sigma }{\delta L^{a}}\frac{\delta \Sigma }{\delta c^{a}}+\frac{%
\delta \Sigma }{\delta L^{i}}\frac{\delta \Sigma }{\delta
c^{i}}+b^{a}\frac{\delta \Sigma }{\delta \overline{c}^{a}}+b^{i}%
\frac{\delta \Sigma }{\delta \overline{c}^{i}}+J\frac{\delta
\Sigma }{\delta \lambda }\right) \;.  \label{ste}
\end{eqnarray}
\item  the diagonal ghost equation \cite{Fazio:2001rm}
\begin{equation}
\mathcal{G}^{i}\Sigma =\Delta _{\mathrm{cl}}^{i}\;,  \label{dg}
\end{equation}
where
\begin{equation}
\mathcal{G}^{i}=\frac{\delta }{\delta c^{i}}+gf^{abi}\overline{c}^{a}\frac{%
\delta \Sigma }{\delta b^{b}}\;,  \label{ghop}
\end{equation}
and
\begin{equation}
\Delta _{\mathrm{cl}}^{i}=-\partial^{2}
\overline{c}^{i}+gf^{abi}\Omega ^{\mu a}A_{\mu }^{b}-\partial
_{\mu }\Omega ^{\mu i }-gf^{abi}L^{a}c^{b}\;.  \label{cl}
\end{equation}
Notice that expression $\left( \ref{cl}\right) $, being linear in
the quantum fields, is a classical breaking.
\item  the diagonal gauge fixing and anti-ghost equations
\begin{eqnarray}
\frac{\delta \Sigma }{\delta b^{i}} &=&\partial_\mu A^{\mu i}\;,  \label{di} \\
\frac{\delta \Sigma }{\delta \overline{c}^{i}}+\partial ^{\mu
}\frac{\delta \Sigma }{\delta \Omega ^{\mu i}} &=&0\;.  \nonumber
\end{eqnarray}
\item  the integrated $\lambda $-equation
\begin{equation}
\int d^{4}x\left( \frac{\delta \Sigma }{\delta \lambda
}+c^{a}\frac{\delta \Sigma }{\delta b^{a}}\right) =0\;, \label{il}
\end{equation}
 expressing in a functional form the on-shell BRST
invariance of the gluon-ghost operator
$\mathcal{O}_{\mathrm{MAG}}$.
\item the diagonal $U(1)^{N-1}$ Ward
identity
\begin{equation}
\mathcal{W}^{i}\Sigma =-\partial ^{2}b^{i}\;,  \label{wi}
\end{equation}
with
\begin{equation}
\;\mathcal{W}^{i}=\partial _{\mu }\frac{\delta }{\delta A_{\mu }^{i}}%
+gf^{abi}\left( A_{\mu }^{a}\frac{\delta }{\delta A_{\mu }^{b}}+c^{a}\frac{%
\delta }{\delta c^{b}}+b^{a}\frac{\delta }{\delta b^{b}}+\overline{c}^{a}%
\frac{\delta }{\delta \overline{c}^{b}}+\Omega ^{\mu
a}\frac{\delta }{\delta \Omega ^{\mu b}}+L^{a}\frac{\delta \Sigma
}{\delta L^{b}}\right) \;. \label{gh-u1}
\end{equation}
This identity follows from the diagonal ghost equation $\left( \ref{dg}%
\right) $ and the Slavnov-Taylor identity $\left( \ref{st}\right)
$.
\end{itemize}
In order to find the foregoing Ward identities, use has been made
of the Jacobi identities (\ref{jacobi2}).
\subsection{Algebraic characterization of the most general local counterterm.}
We mention that all the classical Ward identities of the previous
section can be extended to all orders of perturbation theory
without encountering anomalies. In principle, this can be proven
by means of the algebraic setup of \cite{book} and  of the general
results on the BRST cohomology of gauge theories
\cite{Barnich:2000zw}. It can be understood  in a simple way by
observing that pure Yang-Mills theory in the MAG can be
regularized in a gauge invariant way by employing dimensional
regularization.

In order to characterize the most general invariant counterterm
which can be freely added to all orders of perturbation theory, we
perturb the classical action $\Sigma $ by adding an arbitrary
integrated local polynomial $\Sigma ^{\mathrm{count}}$ in the
fields and external sources of dimension bounded by four and with
zero ghost number, and we require that the perturbed action
$\left( \Sigma +\eta\Sigma ^{\mathrm{count}}\right) $ satisfies
the same Ward identities as $\Sigma $ to the first order in the
perturbation parameter $\eta$, i.e.,
\begin{eqnarray}
\mathcal{S}(\Sigma +\eta \Sigma ^{\mathrm{count}})
&=&0\;+O(\eta ^{2})\;,\nonumber  \label{cds} \\
\mathcal{G}^{i}(\Sigma +\eta \Sigma ^{\mathrm{count}}) &=&\Delta _{%
\mathrm{cl}}^{i}\;+O(\eta ^{2})\;,  \nonumber \\
\frac{\delta (\Sigma +\eta \Sigma ^{\mathrm{count}})}{\delta
b^{i}}
&=&\partial_\mu A^{\mu i}\;+O(\eta^2)\;,  \nonumber \\
\left( \frac{\delta }{\delta \overline{c}^{i}}+\partial ^{\mu
}\frac{\delta
}{\delta \Omega ^{\mu i}}\right) (\Sigma +\eta \Sigma ^{\mathrm{count}%
}) &=&0\;+O(\eta^2)\;,  \nonumber \\
\int d^{4}x\left( \frac{\delta }{\delta \lambda
}+c^{a}\frac{\delta }{\delta b^{a}}\right) (\Sigma +\eta \Sigma
^{\mathrm{count}})
&=&0\;+O(\eta ^{2})\;,  \nonumber \\
\mathcal{W}^{i}(\Sigma +\eta\Sigma ^{\mathrm{count}}) &=&-\partial
^{2}b^{i}\;+O(\eta ^{2})\;.
\end{eqnarray}
This amounts to imposing the following conditions on $\Sigma
^{\mathrm{count}}$
\begin{equation}
\mathcal{B}_{\Sigma }\Sigma ^{\mathrm{count}}=0\;,  \label{bc}
\end{equation}
where $\mathcal{B}_{\Sigma }$ denotes the nilpotent linearized
operator
\begin{equation}
\mathcal{B}_{\Sigma }\mathcal{B}_{\Sigma }=0\;,  \label{np}
\end{equation}
\begin{eqnarray}
\mathcal{B}_{\Sigma } &=&\int d^{4}x\left( \frac{\delta \Sigma
}{\delta
\Omega ^{\mu a}}\frac{\delta }{\delta A_{\mu }^{a}}+\frac{\delta \Sigma }{%
\delta A_{\mu }^{a}}\frac{\delta }{\delta \Omega ^{\mu a
}}+\frac{\delta
\Sigma }{\delta \Omega ^{\mu i}}\frac{\delta }{\delta A_{\mu }^{i}}+\frac{%
\delta \Sigma }{\delta A_{\mu }^{i}}\frac{\delta }{\delta \Omega ^{\mu i}}+%
\frac{\delta \Sigma }{\delta L^{a}}\frac{\delta }{\delta
c^{a}}\right.
\nonumber \\
&+&\left. \frac{\delta \Sigma }{\delta c^{a}}\frac{\delta }{\delta L^{a}}+%
\frac{\delta \Sigma }{\delta L^{i}}\frac{\delta }{\delta
c^{i}}+\frac{\delta
\Sigma }{\delta c^{i}}\frac{\delta }{\delta L^{i}}+b^{a}\frac{\delta }{%
\delta \overline{c}^{a}}+b^{i}\frac{\delta }{\delta \overline{c}^{i}}+J\frac{%
\delta }{\delta \lambda }\right) \;,  \nonumber \\
&&  \label{lb}
\end{eqnarray}
and
\begin{eqnarray}
\mathcal{G}^{i}\Sigma ^{\mathrm{count}} &=&0\;,  \nonumber \\
\frac{\delta \Sigma ^{\mathrm{count}}}{\delta b^{i}} &=&0\;,  \nonumber \\
\left( \frac{\delta }{\delta \overline{c}^{i}}+\partial ^{\mu
}\frac{\delta }{\delta \Omega ^{\mu i}}\right) \Sigma
^{\mathrm{count}} &=&0\;,  \nonumber
\\
\int d^{4}x\left( \frac{\delta }{\delta \lambda
}+c^{a}\frac{\delta }{\delta b^{a}}\right) \Sigma +\varepsilon
\Sigma ^{\mathrm{count}} &=&0\;\;,
\nonumber \\
\mathcal{W}^{i}\Sigma ^{\mathrm{count}} &=&0\;.  \label{rw}
\end{eqnarray}
>From the conditions $\left( \ref{bc}\right) $ and $\left(
\ref{rw}\right) $, it turns out that the most general invariant
counterterm can be written as
\begin{equation}
\Sigma ^{\mathrm{count}}=\frac{-a_0}{4}\int d^{4}x\left( F_{\mu
\nu }^{a}F^{\mu\nu a}+F_{\mu \nu }^{i}F^{\mu \nu i}\right)
+\mathcal{B}_{\Sigma }\Delta ^{-1}\;,  \label{gc}
\end{equation}
where $\Delta ^{-1}$ is an integrated local polynomial with ghost
number $-1$, given by
\begin{eqnarray}
\Delta ^{-1} &=&\int d^{4}x\left( a_{1}L^{a}c^{a}+a_{3}\Omega
^{\mu a}A_{\mu }^{a}+a_{5}\overline{c}^{a}\left(
b^{a}-gf^{abi}\overline{c}^{b}c^{i}\right)
+a_{6}\overline{c}^{a}D_{\mu }^{ab}A^{\mu b
}-\frac{a_{5}}{2}gf^{abc}\overline{c}^{a}\overline{c}^{b}c^{c}\right.\nonumber\\&+&\left.a_{1}\lambda
\left( \frac{1}{2}A_{\mu }^{a}A^{\mu a}+\alpha \overline{c}%
^{a}c^{a}\right)   +\frac{a_{6}}{2}\lambda A_{\mu }^{a}A^{\mu
a}+2\alpha a_{5}\lambda \overline{c}^{a}c^{a}+\frac{a_{13}\zeta
}{2}\lambda J\right) \;. \label{dct}
\end{eqnarray}
We see thus that $\Sigma ^{\mathrm{count}}$ contains six free
independent parameters, namely $a_0$, $a_{1}$, $a_{3}$, $a_{5}$,
$a_{6}$ and $a_{13}$. These parameters can be reabsorbed by means
of a multiplicative renormalization of the gauge coupling constant
$g$, of the gauge and LCO
parameters $\alpha $, $\zeta $, and of the fields $\phi =(A^{\mu a}$, $%
A^{\mu i}$, $c^{a}$, $\overline{c}^{a}$, $c^{i}$, $\overline{c}^{i}$, $%
b^{a}$, $b^{i})$ and sources $\Phi =(\Omega ^{\mu a}$, $\Omega ^{\mu i}$, $%
L^{a}$, $L^{i}$, $\lambda $, $J)$, according to
\begin{equation}
\Sigma (g,\alpha ,\zeta ,\phi ,\Phi )+\eta \Sigma ^{\mathrm{count}%
}=\Sigma (g_{0},\alpha _{0},\zeta _{0},\phi _{0},\Phi _{0})+O(\eta
^{2})\;,  \label{reab}
\end{equation}
with
\begin{equation}
g_{0}=Z_{g}g\;,\;\;\;\;\alpha _{0}=Z_{\alpha }\alpha
\;,\;\;\;\;\zeta _{0}=Z_{\zeta }\zeta \;,  \nonumber
\end{equation}
\begin{eqnarray}
A_{0}^{\mu a} &=&\widetilde{Z}_{A}^{1/2}A^{\mu
a}\;,\;\;\;\;\;\;\;\;\;\;\;A_{0}^{\mu i} =Z_{g}^{-1}A^{\mu i}\;,
\end{eqnarray}
\begin{eqnarray}
c_{0}^{a} =\widetilde{Z}_{c}^{1/2}c^{a}\;,\;\;\;\;\;\;\;\;\;\;
\overline{c}_{0}^{a} =\widetilde{Z}_{c}^{1/2}\overline{c}^{a}\;,
\end{eqnarray}
\begin{eqnarray}
c_{0}^{i}
=Z_{c}^{1/2}c^{i}\;,\;\;\;\;\;\;\;\;\;\;\overline{c}_{0}^{i}
=Z_{c}^{-1/2}\overline{c}^{i}\;,
\end{eqnarray}
\begin{eqnarray}
b_{0}^{a}
=Z_{g}Z_{c}^{1/2}\widetilde{Z}_{c}^{1/2}b^{a}\;,\;\;\;\;\;\;\;\;\;\;
b_{0}^{i} =Z_{g}b^{i}\;,
\end{eqnarray}
\begin{eqnarray}
\Omega _{0}^{\mu a}
=\widetilde{Z}_{A}^{-1/2}Z_{g}^{-1}Z_{c}^{-1/2}\Omega ^{\mu a
}\;,\;\;\;\;\;\;\;\;\;\;\Omega _{0}^{\mu i} =Z_{c}^{-1/2}\;\Omega
^{\mu i}\;,
\end{eqnarray}
\begin{eqnarray}
L_{0}^{a}
=Z_{g}^{-1}\widetilde{Z}_{c}^{-1/2}Z_{c}^{-1/2}L^{a}\;,\;\;\;\;\;\;\;\;\;\;
L_{0}^{i} =Z_{g}^{-1}Z_{c}^{-1}L^{i}\;,
\end{eqnarray}
and
\begin{eqnarray}
J_{0} &=&Z_{L^{a}}^{-2}\widetilde{Z}_{c}^{-1}J=Z_{g}^{2}Z_{c}J\;,
\label{z2}\nonumber\\ \lambda _{0}
&=&Z_{L^{a}}^{-1}\widetilde{Z}_{c}^{-1/2}\lambda
=Z_{g}Z_{c}^{1/2}\lambda \;,
\end{eqnarray}
with
\begin{eqnarray}
Z_{g} &=&1-\eta \frac{a_0 }{2}\;,\nonumber  \label{zf} \\
Z_{\alpha } &=&1+\eta \left( \frac{2a_{5}}{\alpha }+a_0
-2a_{6}\right) \;,  \nonumber \\
Z_{\zeta } &=&1+\eta \left( a_{13}+2a_0 -2a_{1}-2a_{6}\right) \;,
\nonumber \\
\widetilde{Z}_{A}^{1/2} &=&1+\eta \left(\frac{a_0 }{2}%
+a_{3}\right) \;,  \nonumber \\
\widetilde{Z}_{c}^{1/2} &=&1+\frac{\eta }{2}(a_{6}-a_{1})\;,
\nonumber \\
Z_{c}^{1/2} &=&1+\frac{\eta }{2}(a_{6}+a_{1})\;.
\end{eqnarray}
In particular, from eq.$\left( \ref{z2}\right)
$,$\,$ one sees that the renormalization of the source $J$, and thus of the composite operator $\mathcal{O}_{%
\mathrm{MAG}}$, can be expressed in terms of the renormalization
of gauge coupling constant and of the diagonal ghost. This
property follows from the diagonal ghost equation $\left(
\ref{dg}\right) $ and from the the integrated $\lambda $-equation
$\left( \ref{il}\right) $. In particular, for the anomalous
dimension of the gluon-ghost operator
$\mathcal{O}_{\mathrm{MAG}}$, we obtain \cite{Dudal:2003pe}
\begin{equation}
\gamma_{\mathcal{O}_{\mathrm{MAG}}}(g^2)=\mu \frac{\partial }{\partial \mu }%
\log \left( Z_{g}^{2}Z_{c}\right) =-2\left(\frac{\beta(g^2)
}{2g^2}-\gamma _{c^{i}}(g^2)\right), \label{go}
\end{equation}
with
\begin{eqnarray}
\beta(g^2)&=&\mu\frac{\p g^2}{\p\mu}=-g^2\mu\frac{\p}{\p\mu}\ln
Z_g^2\;,\nonumber\\
\gamma_{c^i}(g^2)&=& \mu\frac{\p}{\p\mu}\ln Z_c^{1/2}\;.
\end{eqnarray}
\subsection{The effective potential.}
We  present here the main steps in the construction of the
effective potential for a local composite operator. A more
detailed account of the LCO formalism can be found in
\cite{Verschelde:jj,Knecht:2001cc}.

To obtain the effective potential for the condensate
$\left\langle\mathcal{O}_{\mathrm{MAG}}\right\rangle$, we set the
sources $\Omega_\mu^i$, $\Omega_\mu^a$, $L^a$, $L^i$ and $\lambda$
to zero and consider the renormalized generating functional
\begin{eqnarray}
\label{d1}    \exp(-i\mw(J))&=&\int [D\varphi]\exp iS(J)\;,\nonumber\\
\label{d2}
    S(J)&=&S_{\mathrm{YM}}+S_{\mathrm{MAG}}+S_{\mathrm{diag}}+S_{\mathrm{count}}+\int d^{4}x
\left(Z_JJ\left(\frac{1}{2}\widetilde{Z}_A A_\mu^a A^{\mu
a}+Z_\al\widetilde{Z}_c\al \occ^a
c^a\right)+(\zeta+\delta\zeta)\frac{J^2}{2}\right)\;,
\end{eqnarray}
where $\varphi$ denotes the relevant fields and
$S_{\mathrm{count}}$ is the usual counterterm contribution, i.e.
the part without the composite operator. The quantity
$\delta\zeta$ is the counterterm accounting for the divergences
proportional to $J^2$. Using dimensional regularization throughout
with the convention that $d=4-\varepsilon$, one has the following
identification
\begin{equation}\label{rge5}
    \zeta_0J_0^2=\mu^{-\varepsilon}(\zeta+\delta\zeta)J^2\;.
\end{equation}
The  functional $\mw(J)$ obeys the renormalization group equation
(RGE)
\begin{equation}\label{rge3}
    \left(\mu\frac{\p}{\p\mu}+\beta(g^2)\frac{\p}{\p
g^2}+\al\gamma_\al(g^2)\frac{\p}{\p\al}-\gamma_{\mathcal{O}_{\mathrm{MAG}}}(g^2)\int
d^4x J\frac{\delta}{\delta
    J}+\eta(g^2,\zeta)\frac{\p}{\p\zeta}\right)\mw(J)=0\;,
\end{equation}
where
\begin{eqnarray}\label{rge4}
    \gamma_{\al}(g^2)&=&\mu\frac{\p}{\p\mu}\ln\al=\mu\frac{\p}{\p\mu}\ln
    Z_\al^{-1}\;,\nonumber\\
\eta(g^2,\zeta)&=&\mu\frac{\p}{\p\mu}\zeta\;.
\end{eqnarray}
>From eq.(\ref{rge5}), one finds
\begin{equation}\label{rge6}
\eta(g^2,\zeta)=2\gamma_{\mathcal{O}_{\mathrm{MAG}}}(g^2)\zeta+\delta
(g^2,\al)\;,
\end{equation}
with
\begin{equation}  \label{rge7}
\delta(g^{2},\alpha)=\left(\varepsilon+2\gamma_{\mathcal{O}_{\mathrm{MAG}}}(g^{2})-\beta
(g^{2})\frac{\partial }{\partial g^{2}}%
-\alpha\gamma_{\alpha}(g^{2})\frac{\partial}{\partial\alpha}%
\right)\delta\zeta\;.
\end{equation}
Up to now, the LCO parameter $\zeta$ is still an arbitrary
coupling. As explained in \cite{Verschelde:jj,Knecht:2001cc},
simply setting $\zeta=0$ would give rise to an inhomogeneous RGE
for $\mw(J)$
\begin{equation}  \label{rge8}
    \left(\mu\frac{\p}{\p\mu}+\beta(g^2)\frac{\p}{\p
g^2}+\al\gamma_\al(g^2)\frac{\p}{\p\al}-\gamma_{\mathcal{O}_{\mathrm{MAG}}}(g^2)\int
d^4x J\frac{\delta}{\delta
    J}\right)\mw(J)=\delta(g^2,\al)\int d^4x \frac{J^2}{2}\;,
\end{equation}
and a non-linear RGE for the associated effective action $\Gamma$
for the composite operator $\mathcal{O}_{\mathrm{MAG}}$.
Furthermore, multiplicative renormalizability is lost and by
varying the value of $\delta\zeta$, minima of the effective action
can change into maxima or can get lost. However, $\zeta$ can be
made such a function of $g^2$ and $\al$ so that, if $g^2$ runs
according to $\beta(g^2)$ and $\al$ according to
$\gamma_\al(g^2)$, $\zeta(g^2,\al)$ will run according to its RGE
(\ref{rge6}). This is accomplished by setting $\zeta$ equal to the
solution of the differential equation
\begin{equation}\label{rge9}
\left(\beta(g^2)\frac{\partial}{\partial
g^2}+\al\gamma_\al(g^2,\al)\frac{\partial}{\partial\al}\right)\zeta(g^2,
\al)=2\gamma_{\mathcal{O}_{\mathrm{MAG}}}(g^2)\zeta(g^2,\alpha)+\delta(g^2,\al)\;.
\end{equation}
Doing so, $\mw(J)$ obeys the homogeneous renormalization group
equation
\begin{equation}  \label{rge10}
    \left(\mu\frac{\p}{\p\mu}+\beta(g^2)\frac{\p}{\p
g^2}+\al\gamma_\al(g^2)\frac{\p}{\p\al}-\gamma_{\mathcal{O}_{\mathrm{MAG}}}(g^2)\int
d^4x J\frac{\delta}{\delta J}\right)\mw(J)=0\;.
\end{equation}
To lighten the notation, we will drop the renormalization factors
from now on. One will notice that there are terms quadratic in the
source $J$ present in $\mathcal{W}(J)$, obscuring the usual energy
interpretation. This can be cured  by removing the terms
proportional to $J^2$ in the action to get a generating functional
that is linear in the source, a goal easily achieved by inserting
the following unity,
\begin{equation}  \label{rge11}
1=\frac{1}{N}\int [D\sigma]\exp\left[i\int
d^{4}x\left(-\frac{1}{2\zeta}\left(%
\frac{\sigma}{g}- \mathcal{O}_{\mathrm{MAG}}-\zeta
J\right)^{2}\right)\right]\;,
\end{equation}
with $N$ the appropriate normalization factor, in eq.(\ref{d1}) to
arrive at the Lagrangian
\begin{eqnarray}
\mathcal{L}(A_\mu,\sigma)=-\frac{1}{4}F_{\mu\nu}^{a}F^{\mu\nu
a}-\frac{1}{4}F_{\mu\nu}^{i}F^{\mu\nu
i}+\mathcal{L}_{\textrm{MAG}}+\mathcal{L}_{\textrm{diag}}-\frac{\sigma^2}{2g^2\zeta}
+\frac{1}{g^2\zeta}g\sigma\mathcal{O}_{\mathrm{MAG}}-\frac{1}{2\zeta}\left(\mathcal{O}_{\mathrm{MAG}}\right)^2\;,
\label{rge12}
\end{eqnarray}
while
\begin{eqnarray}
\label{rge13}\exp(-i\mw(J))&=&\int [D\varphi]\exp iS_\sigma(J)\;,\\
\label{rge13bis}S_\sigma(J)&=&\int
d^4x\left(\mathcal{L}(A_\mu,\sigma)+J\frac{\sigma}{g}\right)\;.
\end{eqnarray}
>From eqs.(\ref{d1}) and (\ref{rge13}), one has the following
simple relation
\begin{equation}\label{rge14}
    \left.\frac{\delta\mw(J)}{\delta
J}\right|_{J=0}=-\left\langle\mathcal{O}_{\mathrm{MAG}}\right\rangle=-\left\langle
\frac{\sigma}{g}\right\rangle\;,
\end{equation}
meaning that the condensate $\left\langle
\mathcal{O}_{\mathrm{MAG}}\right\rangle$ is directly related to the
expectation value of the field $\s$, evaluated with the action
$S_\s=\int d^4x\mathcal{L}(A_\mu,\s)$. As it is obvious from
eq.(\ref{rge12}), $\left\langle\sigma\right\rangle\neq0$ is
sufficient to have a tree level dynamical mass for the off-diagonal
fields. At lowest order (i.e. tree level), one finds
\begin{eqnarray}
m_{\mathrm{gluon}}^{\mathrm{off-diag.}}&=&\sqrt{\frac{g\sigma}{\zeta_0}}\;,
\nonumber\\
m_{\mathrm{ghost}}^{\mathrm{off-diag.}}&=&\sqrt{\al
\frac{g\sigma}{\zeta_0}}\;.
\end{eqnarray}
Meanwhile, the diagonal degrees of freedom remain massless. This
could have been established already from the  local $U(1)^{N-1}$
Ward identity (\ref{wi}).
\section{Gauge parameter independence of the vacuum energy.}
We begin this section with a few remarks on the determination of
$\zeta(g^2,\al)$. From explicit calculations in perturbation
theory, it will become clear \footnote{See section V.} that the
RGE functions showing up in the differential equation (\ref{rge9})
look like
\begin{eqnarray}\label{ga1}
    \beta(g^2)&=&-\varepsilon
    g^2-2\left(\beta_0g^2+\beta_1g^2+\cdots\right)\;,\nonumber\\
    \gamma_{\mathcal{O}_{\mathrm{MAG}}}(g^2)&=&\gamma_0(\al)g^2+\gamma_1(\al)g^4+\cdots\;,\nonumber\\
    \gamma_{\al}(g^2)&=&a_0(\al)g^2+a_1(\al)g^4+\cdots\;,\nonumber\\
    \delta(g^2,\al)&=&\delta_0(\al)+\delta_1(\al)g^2+\cdots\;.
\end{eqnarray}
As such, eq.(\ref{rge9}) can be solved  by expanding
$\zeta(g^2,\al)$ in a Laurent series in $g^2$,
\begin{equation}\label{ga2}
    \zeta(g^2,\al)=\frac{\zeta_0(\al)}{g^2}+\zeta_1(\al)+\zeta_2(\al)g^2+\cdots\;.
\end{equation}
More precisely, for the first coefficients $\zeta_0$, $\zeta_1$ of
the expression (\ref{ga2}), one obtains
\begin{eqnarray}
\label{rge21a}2\beta_0\zeta_0+\al
a_0\frac{\partial\zeta_0}{\partial\al}&=&2\gamma_0\zeta_0+\delta_0\;,\nonumber\\
\label{rge21b}2\beta_1\zeta_0+\al
a_0\frac{\partial\zeta_1}{\partial\al}+\al
a_1\frac{\partial\zeta_0}{\partial\al}&=&2\gamma_0\zeta_1+2\gamma_1\zeta_0
+\delta_1\;.\nonumber\\
\end{eqnarray}
Notice that, in order to construct the $n$-loop effective
potential, knowledge of the $(n+1)$-loop RGE functions is needed.

The effective potential calculated with the Lagrangian
(\ref{rge12}) will explicitly depend on the gauge parameter $\al$.
The question arises concerning the vacuum energy $\Evac$, (i.e.
the effective potential evaluated at its minimum); will it be
independent of the choice of $\al$? Also, as it can be seen from
the equations (\ref{rge21a}), each $\zeta_i(\al)$ is determined
through a first order differential equation in $\al$. Firstly, one
has to solve for $\zeta_0(\al)$. This will introduce one arbitrary
integration constant $C_0$. Using the obtained value for
$\zeta_0(\al)$, one can consequently solve the first order
differential equation for $\zeta_1(\al)$. This will introduce a
second integration constant $C_1$, etc. In principle, it is
possible that these arbitrary constants influence the vacuum
energy,  which would represent an unpleasant feature. Notice that
the differential equations in $\al$ for the $\zeta_i$ are due to
the  running of $\al$ in eq.(\ref{rge9}), encoded in the
renormalization group function $\gamma_\al(g^2)$. Assume that we
would have already shown that $\Evac$ does not depend on the
choice of $\al$. If we then set $\al=\al^*$, with $\al^*$ a fixed
point of the RGE for $\al$ at the considered order of perturbation
theory, then equation (\ref{rge9}) determining $\zeta$ simplifies
to
\begin{equation}\label{rge9bis}
\beta(g^2)\frac{\partial}{\partial g^2}\zeta(g^2,
\al^*)=2\gamma_{\mathcal{O}_{\mathrm{MAG}}}(g^2)\zeta(g^2,\alpha^*)+\delta(g^2,\al^*)\;,
\end{equation}
since
\begin{equation}\label{rge9tris}
    \left.\gamma_\al(g^2)\al\right|_{\al=\al^*}=0\;.
\end{equation}
This will lead to simple algebraic equations for the
$\zeta_i(\al^*)$. Hence, no integration constants will enter the
final result for the vacuum energy for $\al=\al^*$, and since
$\Evac$ does not depend on $\al$, $\Evac$ will never depend on the
integration constants, even when calculated for a general $\al$.
Hence, we can put them equal to zero from the beginning for
simplicity.

Summarizing, two questions remain. Firstly, we should prove that
the value of $\al$ will not influence the obtained value for
$\Evac$. Secondly, we should show that there exists a fixed point
$\al^*$. We postpone the discussion concerning the second question
to the next section, giving a positive answer to the first one. In
order to do so,  let us reconsider the generating functional
(\ref{rge13}). We have the following identification, ignoring the
overall normalization factors
\begin{eqnarray}\label{rge34}
    \exp(-i\mw(J))=\int [D\varphi]\exp iS_\sigma(J)=\frac{1}{N}\int
    [D\varphi D\sigma]\exp i\left[S(J)+\int
d^{4}x\left(-\frac{1}{2\zeta}\left(%
\frac{\sigma}{g}-\mathcal{O}_{\mathrm{MAG}}-\zeta
J\right)^{2}\right)\right]\;,
\end{eqnarray}
where $S(J)$ and $S_\sigma(J)$ are given respectively by
eq.(\ref{d2}), and eq.(\ref{rge13bis}). Obviously,
\begin{equation}
\frac{d}{d\alpha }\frac{1}{N}\int [D\sigma ]\exp \left[ i\int
d^{4}x\left( -%
\frac{1}{2\zeta }\left( \frac{\sigma
}{g}-\mathcal{O}_{\mathrm{MAG}}-\zeta J\right) ^{2}\right) \right]
=\frac{d}{d\alpha }1=0\;, \label{rge35}
\end{equation}
so that
\begin{equation}
\frac{d\mathcal{W}(J)}{d\alpha }=-\left.\left\langle s\int
d^{4}x\os\left(\frac{1}{2}c^a\occ^a\right)
 \right\rangle\right|_{J=0}+\textrm{terms}\propto J\;, \label{rge36}
\end{equation}
which follows directly from
\begin{eqnarray}
\frac{dS(J)}{d\alpha } =s\os\int
d^4x\left(\frac{1}{2}c^a\occ^a\right)+\textrm{terms}\propto J\;.
\label{rge37}
\end{eqnarray}
We see that the first term in the right hand side of (\ref{rge37})
is an exact BRST variation. As such, its vacuum expectation value
vanishes. This is the usual argument to prove the gauge parameter
independence in the BRST framework \cite{book}. Note that no
 local operator  $\mathcal{\hat O}$, with
$s\mathcal{\hat O}=\mathcal{O}_{\mathrm{MAG}}$, exists.
Furthermore, extending the action of the BRST transformation on
the $\sigma$-field by
\begin{equation}\label{rge38}
    s\s=gs\mathcal{O}_{\mathrm{MAG}}=-A^{\mu a}D_\mu^{ab}c^b+\al b^a c^a-\al gf^{abi}\occ^a\occ^bc^i-\frac{\al}{2}gf^{abc}\occ^a c^b c^c
\end{equation}
one can easily check that
\begin{equation}\label{rge39}
    s\int d^{4}x\mathcal{L}(A_\mu,\s)=0\;,
\end{equation}
so that we have a BRST invariant $\sigma$-action. Thus, when we
consider the vacuum, corresponding to $J=0$, only the BRST exact
term in eq.(\ref{rge36}) survives. The effective action $\Gamma$
is related to $\mathcal{W}(J)$ through a Legendre transformation
\begin{equation}
\Gamma \left( \frac{\sigma }{g}\right) =-\mathcal{W}(J)-\int
d^{4}yJ(y)\frac{%
\sigma (y)}{g}\;.  \label{rge40}
\end{equation}
The effective potential $V(\sigma )$ is then defined as
\begin{equation}
-V(\sigma )\int d^{4}x=\Gamma \left( \frac{\sigma }{g}\right)\;.
\label{rge41}
\end{equation}
Let $\sigma _{\mathrm{min}}$ be the solution of
\begin{equation}
\frac{dV(\sigma )}{d\sigma } =0\;. \label{rge42}
\end{equation}
From
\begin{equation}\label{rge42bis}
    \frac{\delta }{\delta \left( \frac{\sigma }{g}\right) }\Gamma
    =-J\;,
\end{equation}
it follows that
\begin{equation}
\sigma =\sigma _{\mathrm{min}}\Rightarrow J=0\;,  \label{rge43}
\end{equation}
and hence, we derive from eqs.(\ref{rge40}) and (\ref{rge41}) that
\begin{equation}
\left. \frac{d}{d\alpha }V(\sigma )\right| _{\sigma =\sigma
_{\mathrm{min}}}\int d^{4}x=\left. \frac{d}{d\alpha
}\mathcal{W}(J)\right| _{J=0}\;.  \label{rge44}
\end{equation}
 Thus, due to eq.(\ref{rge36}),
\begin{equation}
\left. \frac{d}{d\alpha }V(\sigma )\right| _{\sigma =\sigma
_{\mathrm{min}}}=0\;. \label{rge45}
\end{equation}
We conclude that the vacuum energy $\Evac$ should be independent
from the gauge parameter $\al$.

A completely analogous derivation was obtained in the case of the
linear gauge \cite{Dudal:2003by}. Nevertheless, in spite of the
previous argument, explicit results in that case showed that
$\Evac$ did depend on $\al$.  In \cite{Dudal:2003by} it was argued
that this apparent disagreement was due to a mixing of different
orders of perturbation theory. Let us explain this with a simple
example. Let us first notice that a key argument in the previous
analysis is that the source $J=0$ vanishes at the end of the
calculations. In practice, $J=0$ is achieved by solving the gap
equation (\ref{rge42}). Moreover, in a power series expansion in
the coupling constant, the derivative of the effective potential
with respect to $\sigma$ will look like
\begin{equation}\label{rge46}
    \left(v_0+v_1g^2+O(g^4)\right)\sigma\;,
\end{equation}
where we assume that we work up to order $g^2$. The corresponding
gap equation reads
\begin{equation}\label{rge47}
v_0+v_1g^2+O(g^4)=0\;.
\end{equation}
Due to eqs.(\ref{rge41}) and (\ref{rge42bis}), one also has
\begin{equation}\label{rge48}
J=g\left(v_0+v_1g^2+O(g^4)\right)\sigma\;.
\end{equation}
Imposing the gap equation (\ref{rge47}) leads to
\begin{equation}\label{rge49}
J=g\left(0+O(g^4)\right)\sigma\;.
\end{equation}
 However, as it can be immediately checked from
expression (\ref{rge34}), there are several terms proportional to
$J$ in the right-hand side of eq.(\ref{rge36}). For instance, one
of them is given by $\frac{\p\zeta}{\p\al}J^2$. Since
\begin{eqnarray}
\frac{\p\zeta}{\p\al}=\frac{\p\zeta_0}{\p\al}\frac{1}{g^2}+\frac{\p\zeta_1}
{\p\al}+O(g^2)\;.
\end{eqnarray}
we find
\begin{equation}\label{rge50}
\frac{\p\zeta}{\p\al}J^2=\left(\frac{\p\zeta_0}{\p\al}v_0^2+\left(
\frac{\p\zeta_0}{\p\al}2v_0v_1+\frac{\p\zeta_1}{\p\al}v_0^2\right)g^2+O(g^4)
\right)\sigma^2\;.
\end{equation}
Squaring the gap equation (\ref{rge47}),
\begin{equation}\label{rge51}
v_0^2+2v_1v_0g^2+O(g^4)=0\;,
\end{equation}
leads to
\begin{equation}\label{rge52}
\frac{\p\zeta}{\p\al}J^2=\left(\frac{\p\zeta_1}{\p\al}v_0^2g^2+O(g^4)\right)
\sigma^2\;.
\end{equation}
We see that, if one consistently works to the first order, terms
such as $\frac{\p\zeta}{\p\al}J^2$ do not equal zero, although
$J=0$ to that order. Terms like those on the right-hand side of
eq.(\ref{rge52}) are cancelled by terms which are formally of
higher order,  requiring thus a mixing of different orders of
perturbation theory. Of course, this problem would not have
occurred if we were be able to compute the effective potential up
to infinite order, an evidently hopeless task. Nevertheless, in
\cite{Dudal:2003by} we succeeded in finding a suitable
modification of the LCO formalism in order to circumvent this
problem and obtaining a well defined gauge independent vacuum
energy $\Evac$, without the need of working at infinite order.
Instead of the action (\ref{d2}), let us consider the following
action
\begin{eqnarray}
\widetilde{S}(\wj)&=&S_{\mathrm{YM}}+S_{\mathrm{MAG}}+S_{\mathrm{diag}}+\int
d^{4}x\left[
\widetilde{J}\mf(g^2,\al)\mathcal{O}_{\mathrm{MAG}} +\frac{\zeta }{2}%
\mf^2(g^2,\al)\widetilde{J}^{2}\right]\;, \label{rge53}
\end{eqnarray}
 where, for the moment, $\mf(g^2,\al)$ is an
arbitrary function of $\al$ of the form
\begin{equation}\label{rge54}
    \mf(g^2,\al)=1+f_{0}(\al)g^2+f_1(\al)g^4+O(g^6)\;,
\end{equation}
and $\wj$ is now the source. The generating functional becomes
\begin{equation}\label{rge55}
    \exp(-i\mathcal{\widetilde{W}}(\wj))=\int[D\phi]\exp
i\widetilde{S}(\wj)\;.
\end{equation}
Taking the functional derivative of $\mathcal{\widetilde{W}}(\wj)$
with respect to $\wj$, we obtain
\begin{equation}
\left. \frac{\delta \mathcal{\widetilde{W}}(\wj)}{\delta
\wj}\right|
_{\wj=0}=-\mf(g^2,\al)\left\langle\mathcal{O}_{\mathrm{MAG}}\right\rangle\;.
\label{rge56}
\end{equation}
Once more, we insert unity via
\begin{equation}  \label{rge57}
1=\frac{1}{N}\int [D\wsigma]\exp\left[i\int
d^{4}x\left(-\frac{1}{2\zeta}\left(%
\frac{\wsigma}{g\mf(g^2,\al)}-\mathcal{O}_{\mathrm{MAG}}-\zeta
\wj\mf(g^2,\al) \right)^{2}\right)\right]\;,
\end{equation}
to arrive at the following Lagrangian
\begin{eqnarray}\label{rge58}
\mathcal{\widetilde{L}}(A_\mu,\wsigma)=-\frac{1}{4}F_{\mu\nu}^{a}F^{\mu\nu
a}-\frac{1}{4}F_{\mu\nu}^{i}F^{\mu\nu i}
+\mathcal{L}_{\mathrm{MAG}}+\mathcal{L}_{\mathrm{diag}}-\frac{\wsigma^2}{2g^2\mf^2(g^2,\al)\zeta}
    +\frac{1}{g^2\mf(g^2,\al)\zeta}g\wsigma
\mathcal{O}_{\mathrm{MAG}}-\frac{1}{2\zeta}\left(\mathcal{O}_{\mathrm{MAG}}
\right)^2 .
\end{eqnarray}
>From the generating functional
\begin{eqnarray}
    \label{rge59}\exp(-i\mathcal{\widetilde{W}}(\wj))&=&\int[D\phi]\exp iS_{\wsigma}(\wj)\;,\\
\label{rge59bis}S_{\wsigma}(\wj)&=&\int
d^4x\left(\mathcal{L}(A_\mu,\wsigma)+\wj\frac{\wsigma}{g}\right)\;.
\end{eqnarray}
it follows that
\begin{eqnarray}
\left. \frac{\delta \mathcal{\widetilde{W}}(\wj)}{\delta
\wj}\right|
_{\wj=0}=-\left\langle\frac{\wsigma}{g}\right\rangle\Rightarrow\left\langle
\wsigma\right\rangle
=g\mf(g^2,\al)\left\langle\mathcal{O}_{\mathrm{MAG}}\right\rangle\;,
\label{rge60}
\end{eqnarray}
 The renormalizability of the action (\ref{rge13bis})
implies that the action (\ref{rge59bis}) will be renormalizable
too. Notice indeed that both actions are connected through the
transformation
\begin{eqnarray}\label{trans1}
    \widetilde{J}&=&\frac{J}{\mathcal{F}(g^2,\al)}\;.
\end{eqnarray}
The tree level off-diagonal masses are now provided by
\begin{eqnarray}\label{rge61bis}
    m_{\mathrm{gluon}}^{\mathrm{off-diag.}}&=&\sqrt{\frac{g\wsigma}{\zeta_0}}\;,\nonumber\\
    m_{\mathrm{ghost}}^{\mathrm{off-diag.}}&=&\sqrt{\al \frac{g\wsigma}{\zeta_0}}\;,
\end{eqnarray}
while the vacuum configuration is determined by solving
 the gap equation
\begin{equation}\label{rge62}
    \frac{d\widetilde{V}(\wsigma)}{d\wsigma}=0\;,
\end{equation}
with $\widetilde{V}(\wsigma)$ the effective potential. Minimizing
$\widetilde{V}(\wsigma)$ will lead to a vacuum energy $\Evac(\al)$
which will depend on $\al$ and the hitherto undetermined functions
$f_i(\al)$ \footnote{At first order, $\Evac$ will depend on
$f_0(\al)$, at second order on $f_0(\al)$ and $f_1(\al)$, etc.}.
We will determine those functions $f_i(\al)$ by requiring that
$\Evac(\al)$ is $\al$-independent. More precisely, one has
\begin{equation}\label{rge100}
    \frac{d\Evac}{d\al}=0\Rightarrow\textrm{first order differential equations in $\al$ for }f_i(\al)\;.
\end{equation}
Of course, in order to be able to determine the $f_i(\al)$, we need
an initial value for the vacuum energy $\Evac$. This corresponds to
initial conditions for the $f_i(\al)$. In the case of the linear
gauges,  to fix the initial condition we employed the Landau gauge
\cite{Dudal:2003by}, a choice which would also be possible in case
of the Curci-Ferrari gauges, since the Landau gauge belongs to these
classes of gauges. This choice of the Landau gauge can be motivated
by observing that the integrated operator $\int d^4x A_\mu^A A^{\mu
A}$ has a gauge invariant meaning in the Landau gauge, due to the
transversality condition $\partial_\mu A^{\mu A}=0$, namely
\begin{equation}\label{rge101}
    (VT)^{-1}\min_{U\epsilon SU(N)}\int d^4x\left[\left(A_\mu^A\right)^U \left(A^{\mu A}\right)^U\right]
=\int d^4x (A_\mu^A A^{\mu A}) \;\; \textrm{ in the Landau
gauge}\;,
\end{equation}
with the operator on the left hand side of eq.(\ref{rge101}) being
gauge invariant. Moreover, the Landau gauge is also an all-order
fixed point of the RGE for the gauge parameter in case of the
linear and Curci-Ferrari gauges. At first glance, it could seem
that it is not possible anymore to make use of the Landau gauge as
initial condition in the case of the MAG, since the Landau gauge
does not belong to the class of gauges we are currently
considering. Fortunately, we shall be able to prove that we can
use the Landau gauge as initial condition for the MAG too. This
will be the content of the next section.

Before turning our attention to this  task, it is worth noticing
that, if one would work up to infinite order, the expressions
(\ref{rge53}) and (\ref{rge59bis}) can be transformed
\emph{exactly} into those of (\ref{d2}), respectively
(\ref{rge13bis}) by means of eq.(\ref{trans1}) and its associated
transformation
\begin{equation}\label{trans2}
    \widetilde{\sigma}=\mathcal{F}(g^2,\al)\sigma\;,
\end{equation}
so that the effective potentials $\widetilde{V}(\wsigma)$ and
$V(\sigma)$ are \emph{exactly} the same at infinite order, and as
such will give rise to the same, gauge parameter independent,
vacuum energy.
\section{\label{sec6} Interpolating between the MAG and the Landau gauge.}
 In this section we shall introduce a generalized
renormalizable gauge which interpolates between the MAG and the
Landau gauge. This will provide a connection between these two
gauges, allowing us to use the Landau gauge as initial condition.
An example of such a generalized gauge, interpolating between the
Landau and the Coulomb gauge was already presented in
\cite{Baulieu:1998kx}. Moreover, we must realize that in the
present case, we must also interpolate between the composite
operator $\frac{1}{2}A_\mu^A A^{\mu A}$ of the Landau gauge and
the gluon-ghost operator $\mathcal{O}_{\mathrm{MAG}}$ of the MAG.
Although this seems to be a highly complicated assignment, there
is an elegant way to treat it.

Consider again the $SU(N)$ Yang-Mills action with the MAG gauge
fixing (\ref{smn2}). For the residual Abelian gauge freedom, we
impose
\begin{eqnarray}
S_{\mathrm{diag}}^{\prime} &=&\int d^{4}x\left( b^{i}\partial
_{\mu }A^{\mu i}+\overline{c}^{i}\partial
^{2}c^{i}+\occ^i\partial_\mu\left(gf^{iab}A^{\mu
a}c^b\right)+\kappa g f^{iab}A_\mu^a\left(\partial_\mu
c^i\right)\occ^b + \kappa
g^{2}f^{iab}f^{icd}\overline{c}^{a}c^{d}A_{\mu
}^{b}A^{\mu c}\right.  \nonumber \\
&-&\left.\kappa gf^{iab}A_{\mu }^{i}A^{\mu a}(b^{b}-
gf^{jbc}\overline{c}^{c}c^{j})
+\kappa gf^{iab}A^{\mu i}(D_{\mu }^{ac}c^{c})\overline{c}%
^{b}+\kappa g^{2}f^{abi}f^{acd}A_{\mu }^{i}A^{\mu
c}c^{d}\overline{c}^{b}\right) \;, \label{5}
\end{eqnarray}
where $\kappa$ is an additional gauge parameter. The gauge fixing
(\ref{5}) can be rewritten as a BRST exact expression
\begin{eqnarray}
S_{\mathrm{diag}}^{\prime} &=&\int d^{4}x
\left[\left(1-\kappa\right)s\left(\occ^i
\partial_\mu A^{\mu i}\right)+\kappa s\overline{s}\left(\frac{1}{2}A_\mu^i A^{\mu i}\right)\right]\;.\label{6}
\end{eqnarray}
Next, we will introduce the following  generalized mass dimension
two operator,
\begin{equation}\label{7}
    \oo=\frac{1}{2}A_\mu^a A^{\mu a}+\frac{\kappa}{2}A_\mu^i A^{\mu i}+\al \occ^a c^a
    \;,
\end{equation}
by means of
\begin{eqnarray}
S_{\mathrm{LCO}}^\prime &=&s\int d^{4}x\;\left( \lambda \oo +\zeta \frac{\lambda J}{2%
}\right) \;  \label{8} \\
&=&\int d^{4}x\;\left( J\oo +\zeta \frac{J^{2}}{2}-\alpha \lambda
b^{a}c^{a}+\lambda A^{\mu a}D_{\mu }^{ab}c^{b}+ \alpha \lambda
\overline{c}^{a}\left(
gf\,^{abi}c^{b}c^{i}+\frac{g}{2}f\,^{abc}c^{b}c^{c}\right)\right.\nonumber\\&-&\left.\kappa
\lambda c^i\partial_\mu A^{\mu i}+\kappa gf^{iab}\lambda
A_\mu^aA^{\mu i}c^b \vphantom{\frac{\zeta}{2}J^2}\right) \;,
\nonumber
\end{eqnarray}
with $\left( J,\lambda \right)$ a BRST doublet of external
sources,
\begin{equation}
s\lambda =J\;,\;\;\;\;sJ=0\;.  \label{9}
\end{equation}
 As in the case of the gluon-ghost operator
(\ref{ggop}), the generalized operator of eq.(\ref{7}) turns out
to be BRST invariant on-shell, a property which can again be
expressed in a functional way, see eq.(\ref{20}).

Let us take a closer look at the action
\begin{equation}\label{rge200}
    \Sigma^\prime=S_{\mathrm{YM}}+S_{\mathrm{MAG}}+S_{\mathrm{diag}}^\prime+S_{\mathrm{LCO}}^{\prime}+S_{\mathrm{ext}}\;.
\end{equation}
The external source part of  the action, $S_{\mathrm{ext}}$, is
the same as given in eq.(\ref{sexr}).

Also, it can be noticed that, for $\kappa\rightarrow0$, the
generalized local composite operator $\mathcal{O}$ of eq.(\ref{7})
reduces to the composite operator $\mathcal{O}_{\mathrm{MAG}}$ of
the MAG, while the diagonal gauge fixing (\ref{6}) reduces to the
Abelian Landau gauge (\ref{abgf}). Said otherwise, for
$\kappa\rightarrow0$, the action $\Sigma^\prime$ of
eq.(\ref{rge200}) reduces to the one we are actually interested in
and which we have discussed in the previous sections.

Another special case is $\kappa\rightarrow1$,
$\alpha\rightarrow0$. Then the gauge fixing terms of
$\Sigma^\prime$ are
\begin{equation}\label{rge201}
    S_{\mathrm{MAG}}+S_{\mathrm{diag}}^\prime=\int d^4xs\left(-A_\mu^A\partial^\mu \occ^A\right)=
\int d^4x\left(\occ^A\partial^\mu D_\mu^{AB}c^B+b^A\partial^\mu
A_\mu^A\right)\;,
\end{equation}
which is nothing else than the Landau gauge. At the same time, we
also have
\begin{equation}\label{rge202}
\lim_{(\al,\kappa)\rightarrow(0,1)}\mathcal{O}=\frac{1}{2}A_\mu^A
A^{\mu A}\;,
\end{equation}
which is the  pure gluon mass operator of the Landau gauge
\cite{Verschelde:2001ia,Dudal:2002pq}.

>From \cite{Dudal:2002pq}, we already know that the Landau gauge
 with the inclusion of the operator $A_\mu^A A^{\mu
A}$ is renormalizable to all orders of perturbation theory.
 On the other hand, in section II, we have proven the
renormalizability for $\kappa=0$. Before we continue our argument,
let us first prove the renormalizability of
 $\Sigma^\prime$ for general $\al$ and $\kappa\neq0$.
The complete action $\Sigma^\prime$, as given in
eq.(\ref{rge200}), is BRST invariant
\begin{equation}
s\Sigma^\prime =0\;,  \label{13}
\end{equation}
and obeys the following identities
\begin{itemize}
    \item The Slavnov-Taylor identity, provided by
\begin{eqnarray}\label{14}
    \mathcal{S}(\Sigma^\prime)&=&\int d^{4}x\left( \frac{\delta \Sigma^\prime }{\delta
\Omega ^{\mu a}}\frac{\delta \Sigma^\prime }{\delta A_{\mu
}^{a}}+\frac{\delta
\Sigma^\prime }{\delta \Omega ^{\mu i}}\frac{\delta \Sigma^\prime }{\delta A_{\mu }^{i}}+%
\frac{\delta \Sigma^\prime }{\delta L^{a}}\frac{\delta \Sigma^\prime }{\delta c^{a}}+\frac{%
\delta \Sigma^\prime }{\delta L^{i}}\frac{\delta \Sigma^\prime
}{\delta c^{i}}\right.
\nonumber \\
&+&\left. b^{a}\frac{\delta \Sigma^\prime }{\delta \overline{c}^{a}}+b^{i}%
\frac{\delta \Sigma^\prime }{\delta
\overline{c}^{i}}+J\frac{\delta \Sigma^\prime }{\delta \lambda
}\right) =0\;.
\end{eqnarray}
    \item The integrated diagonal ghost equation
    \begin{equation}
\mathcal{G}^{i}\Sigma^\prime =\Delta _{\mathrm{cl}}^{i}\;,
\label{15}
\end{equation}
where
\begin{equation}
\mathcal{G}^{i}=\int d^{4}x\left[\frac{\delta }{\delta c^{i}}+gf^{abi}\overline{c}^{a}\frac{%
\delta }{\delta b^{b}}\right]\;,  \label{16}
\end{equation}
and
\begin{equation}
\Delta _{\mathrm{cl}}^{i}=\int d^{4}x\left[gf^{abi}\Omega ^{\mu
a}A_{\mu }^{b}-gf^{abi}L^{a}c^{b}+\kappa\lambda\partial_\mu A^{\mu
i}\right]\;, \label{17}
\end{equation}
a classical breaking.
\item The diagonal anti-ghost equation
\begin{equation}\label{18}
        \frac{\delta \Sigma^\prime }{\delta \overline{c}^{i}}+\partial ^{\mu
}\frac{\delta \Sigma^\prime }{\delta \Omega ^{\mu i}}=0\;,
\end{equation}
    and
\begin{equation}\label{19}
        \frac{\delta \Sigma^\prime }{\delta b^{i}} =\partial_\mu A^{\mu
        i}\;.
\end{equation}
    \item The integrated generalized $\lambda$-equation
    \begin{equation}\label{20}
    \int d^4x\left[\frac{\delta}{\delta\lambda}+c^a\frac{\delta}{\delta b^a}+\kappa c^i\frac{\delta}{\delta
    b^i}\right]\Sigma^\prime=0\;,
\end{equation}
expressing the on-shell BRST invariance of the operator
$\mathcal{O}$ of eq.(\ref{7}).
\end{itemize}
Also in this case,  these Ward identities extend to the quantum
level. Accordingly, the most general local counterterm
$\Sigma^{\prime\mathrm{count}}$ must obey the following
constraints
\begin{eqnarray}
\label{21a}\mathcal{B}_{\Sigma^\prime}\Sigma^{\prime\mathrm{count}}&=&0\;,\nonumber\\
\label{21b}\mathcal{G}^i\Sigma^{\prime\mathrm{count}}&=&0\;,\nonumber\\
\label{21c}\frac{\delta}{\delta b^i}\Sigma^{\prime\mathrm{count}}&=&0\;,\nonumber\\
\label{21d}\left[\frac{\delta}{\delta \occ^i}+\partial_{\mu}\frac{\delta}{\delta\Omega_\mu^i}\right]\Sigma^{\prime\mathrm{count}}&=&0\;,\nonumber\\
\label{21e}\int d^4
x\left[\frac{\delta}{\delta\lambda}+c^a\frac{\delta}{\delta
b^a}\right]\Sigma^{\prime\mathrm{count}}&=&0\;.
\end{eqnarray}
 where $\mathcal{B}_{\Sigma^\prime}$ denotes the
nilpotent,  $\mathcal{B}_{\Sigma^\prime}
\mathcal{B}_{\Sigma^\prime}=0$, linearized operator
\begin{eqnarray}\label{22}
\mathcal{B}_{\Sigma^\prime } &=&\int d^{4}x\left( \frac{\delta
\Sigma^\prime }{\delta
\Omega ^{a\mu }}\frac{\delta }{\delta A_{\mu }^{a}}+\frac{\delta \Sigma^\prime }{%
\delta A_{\mu }^{a}}\frac{\delta }{\delta \Omega ^{a\mu
}}+\frac{\delta
\Sigma^\prime }{\delta \Omega ^{\mu i}}\frac{\delta }{\delta A_{\mu }^{i}}+\frac{%
\delta \Sigma^\prime }{\delta A_{\mu }^{i}}\frac{\delta }{\delta \Omega ^{\mu i}}+%
\frac{\delta \Sigma^\prime }{\delta L^{a}}\frac{\delta }{\delta
c^{a}}\right.
\nonumber \\
&+&\left. \frac{\delta \Sigma^\prime }{\delta c^{a}}\frac{\delta }{\delta L^{a}}+%
\frac{\delta \Sigma^\prime }{\delta L^{i}}\frac{\delta }{\delta
c^{i}}+\frac{\delta
\Sigma^\prime }{\delta c^{i}}\frac{\delta }{\delta L^{i}}+b^{a}\frac{\delta }{%
\delta \overline{c}^{a}}+b^{i}\frac{\delta }{\delta \overline{c}^{i}}+J\frac{%
\delta }{\delta \lambda }\right) \;.
\end{eqnarray}
>From general results on BRST cohomology \cite{Barnich:2000zw}, we
know that the most general, local counterterm can be written as
\begin{eqnarray}\label{23}
\Sigma^{\prime\mathrm{count}}&=&-\frac{a_0^\prime}{4}\int
d^{4}x\left(F_{\mu\nu}^a F^{\mu\nu a}+F_{\mu\nu}^i F^{\mu\nu
i}\right)+\mathcal{B}_{\Sigma}\Delta^{-1}\;,
\end{eqnarray}
where $\Delta^{-1}$ is an integrated local polynomial of ghost
number $-1$ and dimension 4, given by
\begin{eqnarray}\label{24}
\Delta^{-1}&=&\int
d^{4}x\left[\vphantom{\frac{A}{2}}a_1^\prime\Omega_\mu^a A^{\mu
a}+a_2^\prime\Omega_\mu^i A^{\mu i}+a_3^\prime L^a c^a+a_4^\prime
L^i c^i+a_5^\prime\left(\partial_\mu \occ^a\right)A^{\mu
a}+a_5^{\prime\prime} gf^{abi}\occ^a A_\mu^iA^{\mu
b}\right.\nonumber\\&+&a_6^\prime\left(\partial_\mu
\occ^i\right)A^{\mu i}+\left.a_7^\prime
gf^{aic}\occ^a\occ^ic^c+\frac{a_8^\prime}{2}\al
gf^{abi}\occ^a\occ^bc^i+
\frac{a_9^\prime}{2}\al gf^{abc}\occ^a\occ^bc^c+\al a_{10}^\prime b^a\occ^a\right.\nonumber\\
&+&\left.a_{11}^\prime b^i\occ^i+\frac{a_{12}^\prime}{2}\lambda
A_\mu^aA^{\mu a}+\frac{a_{13}^\prime}{2}\kappa\lambda
A_\mu^iA^{\mu i}+a_{14}^\prime\al\lambda\occ^a
c^a+a_{15}^\prime\al\lambda\occ^i
c^i+\frac{a_{16}^\prime}{2}\zeta\lambda J\right]\;.
\end{eqnarray}
The constraints (\ref{21e}) lead to the relations
\begin{eqnarray}\label{25}
a_7^\prime&=&a_{11}^\prime=a_{15}^\prime=0\;,\nonumber\\
a_6^\prime&=&a_2^\prime\;,\nonumber\\
a_{13}^\prime&=&a_4^\prime-a_2^\prime\;,\nonumber\\
a_{8}^\prime&=&-2a_{10}^\prime\;,\nonumber\\
a_{14}^\prime&=&2a_{10}^\prime+a_3^\prime\;,\nonumber\\
a_{12}^\prime&=&a_{3}^\prime-a_{5}^\prime\;,\nonumber\\
a_{5}^{\prime\prime}&=&a_5^\prime+\kappa\left(a_{13}^\prime-a_{3}^\prime\right)\;,\nonumber\\
a_{9}^\prime&=&-a_{10}^\prime\;,\nonumber\\
a_2^\prime&=&a_4^\prime=0\;\;\;\;\;\;\;\;\textrm{and thus
}a_{13}^\prime=0\;.
\end{eqnarray}
Summarizing
\begin{eqnarray}\label{29}
\Delta^{-1}&=&\int d^{4}x\left[a_1^\prime\Omega_\mu^a A^{\mu
a}+a_3^\prime L^a c^a-a_5^\prime\occ^a D_\mu^{ab}A^{\mu b}-\kappa
a_3^\prime gf^{abi}\occ^a A_\mu^i A^{\mu b}-a_{10}^\prime\al
gf^{abi}\occ^a\occ^bc^i\right.\nonumber\\&-&\left.
\frac{a_{10}^\prime}{2}\al gf^{abc}\occ^a\occ^bc^c+\al
a_{10}^\prime
b^a\occ^a+a_3^\prime\lambda\left(\frac{1}{2}A_\mu^aA^{\mu
a}+\al\occ^a c^a\right)-\frac{a_{5}^\prime}{2}\lambda A_\mu^a
A^{\mu a}+2a_{10}^\prime\al\lambda\occ^a
c^a+\frac{a_{16}^\prime}{2}\zeta\lambda J\right]\;.
\end{eqnarray}
In comparison with the case of the MAG, we see that
$\Sigma^{\prime\mathrm{count}}$ also contains six free independent
parameters, namely $a_0^\prime$, $a_{1}^\prime$, $a_{3}^\prime$,
$a_{5}^\prime$, $a_{10}^\prime$ and $a_{13}^\prime$, despite the
fact that the action $\Sigma^\prime$ contains the extra gauge
parameter $\kappa$. These parameters can be reabsorbed by a
suitable multiplicative renormalization of the gauge coupling
constant $g$, of the gauge and LCO parameters $\alpha $, $\kappa$, $\zeta $, and of the fields $\phi =(A^{\mu a}$, $%
A_{\mu i}$, $c^{a}$, $\overline{c}^{a}$, $c^{i}$, $\overline{c}^{i}$, $%
b^{a}$, $b^{i})$ and sources $\Phi =(\Omega ^{\mu a}$, $\Omega^{\mu i}$, $%
L^{a}$, $L^{i}$, $\lambda$, $J)$, according to
\begin{equation}
\Sigma^\prime (g,\alpha ,\kappa,\zeta ,\phi ,\Phi )+\eta \Sigma ^{\prime\mathrm{count}%
}=\Sigma^\prime (g_{0},\alpha _{0},\kappa_0, \zeta _{0},\phi
_{0},\Phi _{0})+O(\eta ^{2})\;,  \label{reabbis}
\end{equation}
where
\begin{equation}
g_{0}=Z_{g}g\;,\;\;\;\;\alpha _{0}=Z_{\alpha }\alpha
\;,\;\;\;\;\zeta _{0}=Z_{\zeta
}\zeta\;,\;\;\;\;\kappa_0=Z_c^{-1}\widetilde{Z}_A^{-1/2}Z_g^{-1}\kappa
\;,  \nonumber
\end{equation}
\begin{eqnarray}
A_{0}^{\mu a} &=&\widetilde{Z}_{A}^{1/2}A^{\mu
a}\;,\;\;\;\;\;\;\;\;\;\;\;A_{0}^{\mu i} =Z_{g}^{-1}A^{\mu i}\;,
\end{eqnarray}
\begin{eqnarray}
c_{0}^{a} =\widetilde{Z}_{c}^{1/2}c^{a}\;,\;\;\;\;\;\;\;\;\;\;
\overline{c}_{0}^{a} =\widetilde{Z}_{c}^{1/2}\overline{c}^{a}\;,
\end{eqnarray}
\begin{eqnarray}
c_{0}^{i}
=Z_{c}^{1/2}c^{i}\;,\;\;\;\;\;\;\;\;\;\;\overline{c}_{0}^{i}
=Z_{c}^{-1/2}\overline{c}^{i}\;,
\end{eqnarray}
\begin{eqnarray}
b_{0}^{a}
=Z_{g}Z_{c}^{1/2}\widetilde{Z}_{c}^{1/2}b^{a}\;,\;\;\;\;\;\;\;\;\;\;
b_{0}^{i} =Z_{g}b^{i}\;,
\end{eqnarray}
\begin{eqnarray}
\Omega _{0}^{\mu a}
=\widetilde{Z}_{A}^{-1/2}Z_{g}^{-1}Z_{c}^{-1/2}\Omega ^{\mu
a}\;,\;\;\;\;\;\;\;\;\;\;\Omega _{0}^{\mu i} =Z_{c}^{-1/2}\;\Omega
^{i\mu }\;,
\end{eqnarray}
\begin{eqnarray}
L_{0}^{a}
=Z_{g}^{-1}\widetilde{Z}_{c}^{-1/2}Z_{c}^{-1/2}L^{a}\;,\;\;\;\;\;\;\;\;\;\;
L_{0}^{i} =Z_{g}^{-1}Z_{c}^{-1}L^{i}\;,
\end{eqnarray}
\begin{eqnarray}
\label{z2bis}J_{0} =Z_{g}^{2}Z_{c}J\;,\;\;\;\;\;\;\;\;\;\;\lambda
_{0} =Z_{g}Z_{c}^{1/2}\lambda \;,
\end{eqnarray}
with
\begin{eqnarray}
Z_{g} &=&1-\eta \frac{a_0^\prime }{2}\;,\nonumber  \label{zfbis} \\
Z_{\alpha } &=&1+\eta \left(2a_{5}^\prime+a_0^\prime
-a_{8}^\prime\right) \;,  \nonumber \\
Z_{\zeta } &=&1+\eta \left( a_{16}^\prime+2a_0^\prime
+2a_{5}^\prime-2a_{3}^\prime\right) \;,
\nonumber \\
\widetilde{Z}_{A}^{1/2} &=&1+\eta \left( \frac{a_0^\prime }{2}%
+a_{1}^\prime\right) \;,  \nonumber \\
\widetilde{Z}_{c}^{1/2} &=&1-\frac{\eta
}{2}(a_{5}^\prime+a_{3}^\prime)\;,
\nonumber \\
Z_{c}^{1/2} &=&1+\frac{\eta }{2}(a_{3}^\prime-a_{5}^\prime)\;.
\end{eqnarray}
We see thus that the additional gauge parameter $\kappa$ does not
renormalize in an independent way. Furthermore, from
eq.(\ref{z2bis}), we  notice that the relation (\ref{go}) is
generalized to the operator $\mathcal{O}$, i.e.
\begin{equation}\label{gobis}
    \gamma_\mathcal{O}(g^2)=-2\left(\frac{\beta(g^2)
}{2g^2}-\gamma _{c^{i}}(g^2)\right)\;.
\end{equation}
Summarizing, we have constructed a renormalizable gauge that is
labelled by a couple of parameters $(\al,\kappa)$. It allows us to
introduce a  generalized composite operator $\mathcal{O}$, given
by eq.(\ref{7}), which embodies the  local operator $A_\mu^A
A^{\mu A}$ of the Landau gauge as well as the operator
$\mathcal{O}_{\mathrm{MAG}}$ of the MAG. To construct the
effective potential, one sets all sources equal to zero, except
$J$, and introduces unity to remove the $J^2$ terms. A completely
analogous argument as  the one given in section III allows to
conclude that the minimum value of $V(\sigma)$, thus $\Evac$, will
be independent of $\al$ and $\kappa$, essentially because the
derivative with respect to $\al$ as well as with respect to
$\kappa$ is BRST exact, up to terms in the source $J$. This
independence of $\al$ and $\kappa$ is again only assured at
infinite order in perturbation theory, so we can generalize the
construction, proposed in section III, by making the function
$\mf$ of eq.(\ref{rge54}) also dependent on $\kappa$. The
foregoing analysis is sufficient to make sure that we can use the
Landau gauge result for $\Evac$ as the initial condition for the
vacuum energy of the MAG. Moreover, we are now even in the
position to answer the question about the existence of a fixed
point of the RGE for the gauge parameter $\al$, which was
necessary to certify that no arbitrary constants would enter the
results for $\Evac$. We already mentioned that the Landau gauge,
i.e. the case $(\al,\kappa)=(0,1)$, is a renormalizable model
\cite{Dudal:2002pq}, i.e. the Landau gauge is stable against
radiative corrections. This can be reexpressed by saying that
$(\al,\kappa)=(0,1)$ is a fixed point of the RGE for the gauge
parameters, and this to all orders of perturbation theory.
\section{Numerical results for $SU(2)$.}
After a quite lengthy formal construction of the LCO formalism in
the case of the MAG, we are now  ready to present explicit
results. In this paper, we will restrict ourselves to the
evaluation of the one-loop effective potential in the case of
$SU(2)$. As renormalization scheme, we adopt the modified minimal
substraction scheme ($\MSbar$). Let us give here, for further use,
the values of the one-loop anomalous dimensions of the relevant
fields and couplings in the case of $SU(2)$. In our conventions,
one has \cite{Shinohara:2001cw,Ellwanger:2002sj,Kondo:2003sw}
\begin{eqnarray}
\label{ex1}    \gamma_{c^i}(g^2)&=&\left(-3-\al\right)\frac{g^2}{16\pi^2}+O(g^4)\;,\\
\label{ex2}
\gamma_\al(g^2)&=&\left(-2\al+\frac{8}{3}-\frac{6}{\al}\right)\frac{g^2}{16\pi^2}+O(g^4)\;,
\end{eqnarray}
while
\begin{equation}\label{ex3}
    \beta(g^2)=-\varepsilon g^2-2\left(\frac{22}{3}\frac{g^4}{16\pi^2}\right)+O(g^6)\;,
\end{equation}
and exploiting the relation (\ref{go})
\begin{equation}
\gamma _{O_{\mathrm{MAG}}}(g^2)=\left( \frac{26}{3}%
-2\alpha \right)\frac{g^2}{16\pi^2}+O(g^4) \;,  \label{ex4}
\end{equation}
a result consistent with that of \cite{Ellwanger:2002sj}.

The reader will notice that we have given only the 1-loop values
of the anomalous dimensions, despite the fact that we have
announced that one needs $(n+1)$-loop knowledge of the RGE
functions to determine the $n$-loop potential. As we shall see
soon, the introduction of the function $\mathcal{F}(g^2,\al)$ and
the use of the Landau gauge as initial condition allow us to
determine the 1-loop results we are interested in, from the 1-loop
RGE functions only.

Let us first determine the counterterm $\delta\zeta$. For the
generating functional $\mathcal{W}(J)$, we find at 1-loop
\footnote{We will do the transformation of $\mathcal{W}(J)$ to
$\mw(\wj)$ only at the end.}
\begin{eqnarray}\label{ex5}
\mathcal{W}(J)&=&\int
d^dx\left(-\left(\zeta+\delta\zeta\right)\frac{J^2}{2}\right)
+i\ln\det\left[\delta^{ab}\left(\partial^2+\al J\right)\right]
-\frac{i}{2}\ln\det\left[\delta^{ab}\left(\left(\partial^2+J\right)g_{\mu\nu}-\left(1-\frac{1}{\al}\right)
\partial_\mu\partial_\nu\right)\right]\;,\nonumber\\
\end{eqnarray}
and employing
\begin{equation}\label{ex5bis}
   \ln\det\left[\delta^{ab}\left(\left(\partial^2+J\right)g_{\mu\nu}-\left(1-\frac{1}{\al}\right)
\partial_\mu\partial_\nu\right)\right]=\delta^{aa}\left[(d-1)\mathrm{tr}\ln\left(\partial^2+J\right)+\mathrm{tr}\ln\left(\partial^2+\al J\right)\right]\;,
\end{equation}
with
\begin{equation}\label{ex5tris}
\delta^{aa}=N(N-1)=2\textrm{ for }N=2\;,
\end{equation}
one can calculate the divergent part of eq.(\ref{ex5}),
\begin{eqnarray}\label{ex6}
    \mathcal{W}(J)&=&\int d^4x\left[-\delta\zeta\frac{J^2}{2}-\frac{3}{16\pi^2}J^2\frac{1}{\varepsilon}
-\frac{1}{16\pi^2}\al^2
J^2\frac{1}{\varepsilon}+\frac{1}{8\pi^2}\al^2
J^2\frac{1}{\varepsilon}\right]\,.
\end{eqnarray}
Consequently,
\begin{equation}\label{ex7}
    \delta\zeta=\frac{1}{8\pi^2}\left(\al^2-3\right)\frac{1}{\varepsilon}+O(g^2)\;.
\end{equation}
Next, we can compute the RGE function $\delta(g^2,\al)$ from
eq.(\ref{rge7}),  obtaining
\begin{equation}\label{ex8}
    \delta(g^2,\al)=\frac{\al^2-3}{8\pi^2}+O(g^2)\;.
\end{equation}
Having determined this, we are ready to calculate $\zeta_0$. The
differential equation (\ref{rge21a}) is solved by
\begin{equation}\label{ex9}
    \zeta_0(\al)=\al+\left(9-4\al+3\al^2\right)C_0\;,
\end{equation}
with $C_0$ an integration constant. As already explained in the
previous sections, we can  consistently put $C_0=0$. Here, we have
written it explicitly to illustrate that, if $\al$ would
 coincide with the 1-loop fixed point of the RGE for
the gauge parameter, the part proportional to $C_0$ in
eq.(\ref{ex9}) would drop. Indeed, the equations $9-4\al+3\al^2=0$
and $-2\al+\frac{8}{3}-\frac{6}{\al}=0$,  stemming from
eq.(\ref{ex2}), are the same. Moreover, we also notice that this
equation has only complex valued solutions. Therefore, it is even
more important to have made the connection between the MAG and the
Landau gauge by embedding them in a bigger class of gauges, since
then we have the fixed point, even at all orders. In what follows,
it is understood that $\zeta_0=\al$.

We now have all the ingredients to construct the 1-loop effective
potential $\widetilde{V}_1(\wsigma)$,
\begin{eqnarray}\label{ex10}
\widetilde{V}_1(\wsigma)&=&\frac{\wsigma^2}{2\zeta_0}
\left(1-\left(2f_0+\frac{\zeta_1}{\zeta_0}\right)g^2\right)
+i\ln\det\left[\delta^{ab}\left(\partial^2+\al\frac{g\wsigma}{\zeta_0}\right)
\right] \nonumber \\
&& -~\frac{i}{2}\ln\det\left[\delta^{ab}\left(\left(\partial^2
+\frac{g\wsigma}{\zeta_0}\right)g_{\mu\nu}-\left(1-\frac{1}{\al}\right)
\partial_\mu\partial_\nu\right)\right]\;,\nonumber\\
\end{eqnarray}
or, after renormalization
\begin{eqnarray}\label{ex11}
\widetilde{V}_1(\wsigma)&=&\frac{\wsigma^2}{2\zeta_0}\left(1-\left(2f_0+\frac{\zeta_1}{\zeta_0}\right)g^2\right)
+\frac{3}{32\pi^2}\frac{g^2\wsigma^2}{\zeta_0^2}\left(\ln\frac{g\wsigma}{\zeta_0\omu^2}-\frac{5}{6}\right)-
\frac{1}{32\pi^2}\frac{g^2\al^2\wsigma^2}{\zeta_0^2}\left(\ln\frac{g\al\wsigma}{\zeta_0\omu^2}-\frac{3}{2}\right)\;.
\end{eqnarray}
We did not explicitly write the divergences and counterterms in
eq.(\ref{ex12}), since by construction we know that the formalism
is renormalizable, so they would have cancelled amongst
 each other. This can be checked explicitly by using
the unity of (\ref{rge11}) with counterterms included. It can also
be checked explicitly that $\widetilde{V}_1(\wsigma)$ obeys the
renormalization group
\begin{equation}\label{ex12}
    \mu\frac{d}{d\mu}\widetilde{V}_1(\wsigma)=0+\textrm{terms of higher order}\;,
\end{equation}
by using the RGE functions (\ref{ex1})-(\ref{ex4}) and the fact
that the anomalous dimension of $\wsigma$ is given by
\begin{equation}\label{ex13}
\gamma_{\wsigma}(g^2)
=\frac{\beta(g^2)}{2g^2}+\gamma_{\mathcal{O}_{\mathrm{MAG}}}(g^2)
+\mu\frac{\p\ln\mf(g^2,\al)} {\p\mu}\;,
\end{equation}
which is immediately verifiable from eq.(\ref{rge60}).

We now search for the vacuum configuration by minimizing
$\widetilde{V}_1(\wsigma)$ with respect to $\wsigma$. We will put
$\omu^2=\frac{g\wsigma}{\zeta_0}$ to exclude possibly large
logarithms, and find two solutions of the gap equation
\begin{eqnarray}\label{ex13bis}
&&\left.\frac{d\widetilde{V}_1}{d\sigma}\right|_{\omu^2=\frac{g\wsigma}{\zeta_0}}=0\nonumber\\
&\Leftrightarrow&
\frac{\wsigma}{\zeta_0}\left(1-\left(2f_0+\frac{\zeta_1}{\zeta_0}\right)g^2\right)
+\frac{3}{16\pi^2}\frac{g^2\wsigma}{\zeta_0^2}\left(-\frac{5}{6}\right)+\frac{3}{32\pi^2}\frac{g^2\wsigma}{\zeta_0^2}
-\frac{1}{16\pi^2}\frac{g^2\al^2\wsigma}{\zeta_0^2}\left(\ln\al-\frac{3}{2}\right)-\frac{1}{32\pi^2}\frac{g^2\al^2\wsigma}{\zeta_0^2}=0\;,\nonumber\\
\end{eqnarray}
namely
\begin{eqnarray}
    \label{ex14}\wsigma&=&0\;,\\
    \label{ex14bis}y&\equiv&\left.\frac{g^2N}{16\pi^2}\right|_{N=2}=\frac{2\zeta_0}{16\pi^2\left(2f_0\zeta_0+\zeta_1\right)+\al^2\ln\al-\al^2+1}\;.
\end{eqnarray}
The quantity $y$ is the relevant expansion parameter, and should
be sufficiently small to have a sensible expansion.The value for
$\left\langle\wsigma\right\rangle$ corresponding to
eq.(\ref{ex14bis}) can be extracted from the 1-loop coupling
constant
\begin{equation}\label{ex15}
    g^2(\omu)=\frac{1}{\beta_0\ln\frac{\omu^2}{\lms^2}}\;.
\end{equation}
The first solution (\ref{ex14}) corresponds to the usual,
perturbative vacuum ($\Evac=0$), while eq.(\ref{ex14bis}) gives
rise to a dynamically favoured vacuum with energy
\begin{eqnarray}
    \label{ex16}\Evac&=&-~\frac{1}{64\pi^2}\left(3-\al^2\right)
\left(m_{\mathrm{gluon}}^{\mathrm{off-diag}}\right)^4\;,\\
\label{ex16bis}m_{\mathrm{gluon}}^{\mathrm{off-diag}}&=&e^{\frac{3}{22y}}\lms\;.
\end{eqnarray}
Eq.(\ref{ex16}) is obtained upon substitution of
eq.(\ref{ex13bis}) into eq.(\ref{ex11}). From eq.(\ref{ex16}), we
notice that at the 1-loop approximation, $\al^2\leq3$ must be
fulfilled in order to have $\Evac\leq0$. In principle, the unknown
function $f_0(\al)$ can be determined by solving the differential
equation
\begin{eqnarray}\label{ex17}
    \frac{d\Evac}{d\al}=0&\Leftrightarrow&2\al\left(m_{\mathrm{gluon}}^{\mathrm{off-diag}}\right)^4
+4\left(\al^2-3\right)\left(m_{\mathrm{gluon}}^{\mathrm{off-diag}}\right)^3\frac{dm_{\mathrm{gluon}}^{\mathrm{off-diag}}}{d\al}=0\nonumber\\
&\Leftrightarrow& \al+\frac{3-\al^2}{y^2}\left(\frac{\p y}{\p
\al}+\frac{\p y}{\p \zeta_0}\frac{\p \zeta_0}{\p \al}+\frac{\p
y}{\p \zeta_1}\frac{\p \zeta_1}{\p \al}+\frac{\p y}{\p
f_0}\frac{\p f_0}{\p \al }\right)=0
\end{eqnarray}
with initial condition $\Evac(\al)=\Evac^{\mathrm{Landau}}$.
However,  to solve eq.(\ref{ex17}) knowledge of $\zeta_1$ is
needed. Since we are not interested in $f_0(\al)$ itself, but
rather in the value  of the vacuum energy $\Evac$, the
off-diagonal mass $m_{\mathrm{gluon}}^{\mathrm{off-diag}}$ and the
expansion parameter $y$,  there is a more direct way to proceed,
without having to solve the eq.(\ref{ex17}).  Let us first give
the Landau gauge value for $\Evac$ in the case $N=2$, which can be
easily obtained from \cite{Verschelde:2001ia,Browne:2003uv},
\begin{equation}\label{ex18}
    \Evac^{\mathrm{Landau}}=-\frac{9}{128\pi^2}e^{\frac{17}{6}}\lms^4\;.
\end{equation}
Since the construction is such that $\Evac(\al)=
\Evac^{\mathrm{Landau}}$, we can equally well solve
\begin{equation}\label{ex19}
-\frac{9}{128\pi^2}e^{\frac{17}{6}}\lms^4=-\frac{1}{64\pi^2}\left(3-\al^2\right)\left(m_{\mathrm{gluon}}^{\mathrm{off-diag}}\right)^4\;,
\end{equation}
which gives  the lowest order mass
\begin{equation}\label{ex20}
m_{\mathrm{gluon}}^{\mathrm{off-diag}}=\left(\frac{9}{2}\frac{e^{\frac{17}{6}}}{3-\al^2}\right)^\frac{1}{4}\lms\;,
\end{equation}
and hence
\begin{equation}\label{ex20bis}
m_{\mathrm{ghost}}^{\mathrm{off-diag}}=\sqrt\al\left(\frac{9}{2}\frac{e^{\frac{17}{6}}}{3-\al^2}\right)^\frac{1}{4}\lms\;,
\end{equation}
The result (\ref{ex20}) can be used to determine $y$. From
eq.(\ref{ex16bis}) one easily finds
\begin{equation}\label{ex21}
    y=\frac{36}{187+66\ln\frac{9}{2\left(3-\al^2\right)}}\;.
\end{equation}
We see thus that, for the information we are currently interested
in, we do not need explicit knowledge of $\zeta_1$ and $f_0$. We
want to remark that, if $\zeta_1$ were known, the value for $y$
obtained in eq.(\ref{ex21}) can be used to determine $f_0$ from
eq.(\ref{ex14bis}). This is a nice feature, since the possibly
difficult differential equation (\ref{ex17}) never needs to be
solved in this fashion.
\begin{figure}[t]\label{fig1}
\begin{center}
    \scalebox{0.7}{\includegraphics{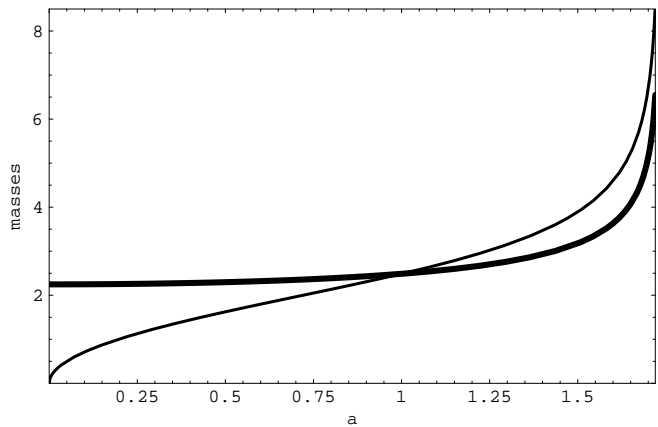}}
    \caption{The off-diagonal gluon (fat line) and ghost mass (thin line) in function of $\alpha$. Masses are in units of $\lms$. }
\end{center}
\end{figure}
\begin{figure}[t]\label{fig2}
\begin{center}
    \scalebox{0.7}{\includegraphics{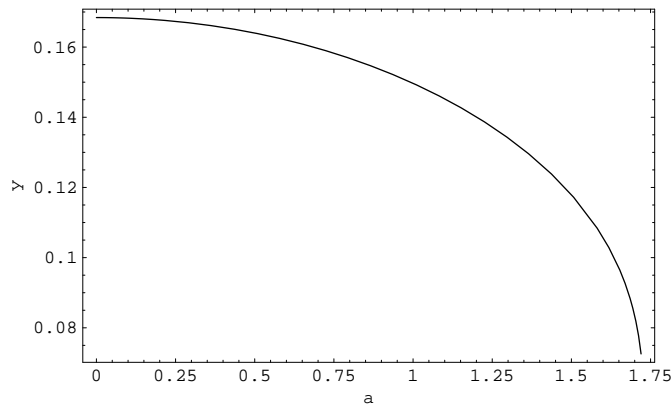}}
    \caption{The expansion parameter $y$ as a function of $\al$.}
\end{center}
\end{figure}
In Fig. 1, we have plotted the off-diagonal gluon mass
(\ref{ex20}) and ghost mass (\ref{ex20bis}) for
$0\leq\al\leq\sqrt{3}$. We notice that the masses grow to $\infty$
for increasing $\al$, while the expansion parameter $y$ drops to
zero, as it is clear from Fig. 2. The relative smallness of $y$
means that our perturbative analysis should give qualitatively
meaningful results. Before we come to the conclusions, let us
consider the limit $\al\rightarrow0$, corresponding to the
``real'' MAG $D_\mu^{ab}A^{\mu b}=0$. One finds
\begin{eqnarray}\label{ex22}
    m_{\mathrm{gluon}}^{\mathrm{off-diag}}&=&\left(\frac{3}{2}e^{\frac{17}{6}}\right)^{\frac{1}{4}}\lms\approx2.25\lms\nonumber\;,\\
    y&=&\frac{36}{187+66\ln\frac{3}{2}}\approx0.168\;.
\end{eqnarray}

\section{Discussion and conclusion.}
The aim of this paper was to give analytic evidence, as expressed by
eq.(\ref{ex22}), of the dynamical mass generation for off-diagonal
gluons in Yang-Mills theory quantized in the maximal Abelian gauge.
This mass can be seen as support for the Abelian dominance
\cite{Ezawa:bf,Suzuki:1989gp,Hioki:1991ai} in that gauge. This
result is in qualitative agreement with the lattice version of the
MAG, were such a mass was also reported
\cite{Amemiya:1998jz,Bornyakov:2003ee}. The off-diagonal lattice
gluon propagator could be fitted by $\frac{1}{p^2+m^2}$, which is in
correspondence with the tree level propagator we find. We have been
able to prove the existence of the off-diagonal mass by
investigating the condensation of a mass dimension two operator,
namely $(\frac{1}{2}A_\mu^a A^{\mu a}+\al \occ^a c^a)$. It was shown
how a meaningful, renormalizable effective potential for this local
composite operator can be constructed.  By evaluating this potential
explicitly at 1-loop order in the case of $SU(2)$, the formation of
the condensate is favoured since it lowers the vacuum energy. The
latter does not depend on the choice of the gauge parameter $\al$,
at least if one would work to infinite order in perturbation theory.
We have explained in short the problem at finite order and discussed
a way to overcome it. Moreover, we have been able to interpolate
between the Landau gauge and the MAG by unifying them in a larger
class of renormalizable gauges. This observation was used to prove
that the vacuum energy of Yang-Mills theory in the MAG due to its
mass dimension two condensate should be the same as the vacuum
energy of Yang-Mills theory in the Landau gauge with the much
explored condensate $\left\langle A_\mu^A A^{\mu A}\right\rangle$.
It is worth noticing that all the gauges, where a dimension two
condensate provides a dynamical gluon mass parameter, such as the
Landau gauge \cite{Verschelde:2001ia}, the Curci-Ferrari gauges
\cite{Dudal:2003gu}, the linear gauges \cite{Dudal:2003by} and the
MAG, can be connected to each other, either directly (e.g.
Landau-MAG) or via the Landau gauge (e.g. MAG and linear gauges).
This also implies that, if $\left\langle A_\mu^A A^{\mu
A}\right\rangle\neq0$ in the Landau gauge, the analogous condensates
in the other gauges cannot vanish either. Then the question arises
if this correspondence between different gauges could be stretched
further to for instance the Coulomb gauge, where the possibility of
a condensate $\left\langle A_i^A A^{i A}\right\rangle$ was already
advocated some time ago in \cite{Greensite:1985vq}.  However, it is
worth remarking that this might be a more complicated task, since
the Coulomb gauge is not a covariant gauge fixing, and as such its
analysis within the algebraic remormalization framework \cite{book}
is not straightforward.

Needless to say, the present work is far from being complete. First
of all, an explicit calculation at two loop order and for general
gauge group $SU(N)$ would be interesting. We also limited our
computations to the tree level order. In principle, one should
evaluate the off-diagonal gluon polarization in order to get further
information on the structure of the propagator. A first step in this
direction was taken in the case of the Landau gauge in
\cite{Browne:2004mk}. It is unknown what will happen at higher
orders in the MAG, but it is likely that the external momentum $Q^2$
will enter through loop corrections and influence the possible
position of a pole in the propagator. The ghost condensation, that
was first investigated in \cite{Schaden:1999ew,Kondo:2000ey} as a
possible mechanism behind the off-diagonal mass, and later on was
shown to be tachyonic \cite{Dudal:2002xe,Sawayanagi:2003dc}, could
enter this polarization too. This would require a more complete
treatment of the ghost condensation in the MAG, along the lines of
\cite{Dudal:2003dp}, where these condensates were considered in more
detail in the case of the Curci-Ferrari and Landau gauge. Another
issue which deserves attention is the behaviour of the diagonal
gluon. In \cite{Bornyakov:2003ee}, it was found that the diagonal
gluon propagator also contain a mass parameters, with
$m_\mathrm{gluon}^{\mathrm{diag}}\approx\frac{1}{2}m_\mathrm{gluon}^{\mathrm{off-diag}}$,
while in \cite{Amemiya:1998jz} the diagonal gluon was reported to
behave like a light or massless particle. For completeness, we
remind that these lattice simulations were both performed in the
case of $SU(2)$. We want to remark that a condensation of the
composite diagonal operator $A_\mu^i A^{\mu i}$ cannot occur within
our approach, since this is forbidden by the diagonal local
$U(1)^{N-1}$ Ward identity (\ref{wi}). In principle, one could add
an extra source term like $\frac{1}{2}\rho A_\mu^i A^{\mu i}$, but
it does not seem possible to prove the renormalizability of this
operator in the MAG. This might be consistent with the result of
\cite{Bornyakov:2003ee}, since the diagonal gluon propagator could
\emph{not} be fitted with a Yukawa propagator $\frac{1}{p^2+m^2}$,
in contrast with the off-diagonal gluon propagator which could be
fitted with $\frac{1}{p^2+m^2}$. This could mean that the diagonal
mass parameter is of a different nature compared to the off-diagonal
one. A possible speculation is that it might have to do with Gribov
copies, since a fit $\frac{p^2}{p^4+m^4}$ did work for the diagonal
propagator \cite{Bornyakov:2003ee}.

Our analysis of the MAG condensate was also restricted to the purely
perturbative level. One could imagine calculating in a certain
non-trivial background. The vacuum energy calculated in one gauge
should still be the same as the one calculated in the other gauge.
In this context, and keeping in mind that monopole condensation is
an essential ingredient of the dual superconductor picture, it might
be worth noticing that the role of $\left\langle A_\mu^A A^{\mu
A}\right\rangle$ as an order parameter for monopole condensation was
investigated in the Landau gauge by the authors of
\cite{Gubarev:2000nz}, based on a similar observation in compact QED
\cite{Gubarev:2000eu}. We note that an off-diagonal gluon mass can
serve as a starting point to derive low energy (dual) Abelian models
for Yang-Mills theories, see for example
\cite{Kondo:2000sh,Deguchi:2003zi,Freire:2001nd}.

Let us conclude with a few considerations on the issues of the
degrees of freedom and of the unitarity when the gluons attain a
dynamical mass, as a consequence of a nonvanishing dimension two
condensate  $\left\langle \mathcal{O}_{\mathrm{MAG}}\right\rangle =
\left\langle \frac{1}{2}A_\mu^a A^{\mu a}+\al \occ^a c^a
\right\rangle $. One possible way to look at the degrees of freedom
associated to a given field is through its propagator. From the pole
of the propagator one gets information about the mass of the field,
while from its residue one learns about polarization states.
However, the propagation of the field has to occur in some vacuum.
In other words, the kind of vacuum in which the field propagates has
to be supplemented. In our case, this task is achieved by the LCO\
Lagrangian, eq.(\ref{rge12}), i.e.
\begin{eqnarray}
\mathcal{L}(A_\mu,\sigma)=-\frac{1}{4}F_{\mu\nu}^{a}F^{\mu\nu
a}-\frac{1}{4}F_{\mu\nu}^{i}F^{\mu\nu
i}+\mathcal{L}_{\textrm{MAG}}+\mathcal{L}_{\textrm{diag}}-\frac{\sigma^2}{2g^2\zeta}
+\frac{1}{g^2\zeta}g\sigma\mathcal{O}_{\mathrm{MAG}}-\frac{1}{2\zeta}\left(\mathcal{O}_{\mathrm{MAG}}\right)^2\;,
\label{f1}
\end{eqnarray}
which allows one to take into account the effects related to having
a nontrivial vacuum corresponding to the nonvanishing dimension two
condensate $\left\langle \mathcal{O}_{\mathrm{MAG}}\right\rangle$,
as expressed by the identity
\begin{equation}
\left\langle \sigma \right\rangle =g \left\langle
\mathcal{O}_{\mathrm{MAG}}\right\rangle \;. \label{f2}
\end{equation}
That this is the preferred vacuum follows from the observation that
the vacuum energy is lowered by the condensate $\left\langle
\mathcal{O}_{\mathrm{MAG}}\right\rangle$. Expanding thus around
$\left\langle \sigma \right\rangle\neq 0$, a dynamical tree level
mass $m_{\mathrm{gluon}}^{\mathrm{off-diag.}}$ for the off-diagonal
gluons is generated in the gauge fixed Lagrangian (\ref{f1}), namely
\begin{equation}
m_{\mathrm{gluon}}^{\mathrm{off-diag.}}=\sqrt{\frac{g\left\langle
\sigma \right\rangle}{\zeta_0}}\;. \label{f3}
\end{equation}
Therefore, in the condensed vacuum, $\left\langle \sigma
\right\rangle\neq 0$, the Lagrangian (\ref{f1}) accounts for
off-diagonal massive gluons. However, we emphasize that this
dynamical mass parameter occurs as the result of a particular
condensate. It is not a free parameter of the gauge fixed theory,
its value being determined by a gap equation. Concerning now the
unitarity of the resulting theory, it should be remarked that, due
to confinement, gluons and quarks are only to be called physical at
a very high energy scale $Q^2$, where they behave almost freely and
asymptotic states can be related to them, thanks to asymptotic
freedom. At very high energies, our dynamically massive action might
be unitary: a renormalization group improvement could induce quantum
corrections such that the mass parameter runs to zero for
$Q^2\rightarrow\infty$. Otherwise said, the corrections induced by
this dynamical mass on the scattering amplitudes are expected to
become less and less important as the energy of the process
increases, so that the amplitudes of the massless case are in fact
recovered. Such a scenario would be analogous to the behaviour of
the dynamical mass parameter discussed by Cornwall in
\cite{Cornwall:1981zr}. For very high $Q^2$, one does indeed expect
that perturbative Yang-Mills theory with massless gluons, having two
physical degrees of freedom, describes the physical spectrum and
that non-perturbative corrections are absent.

To decide if our resulting theory is unitary at smaller $Q^2$, one
should know how to take into account the effects of confinement,
which now cannot be neglected. This would amount to knowing how to
construct out of our Lagrangian (\ref{f1}) the low energy spectrum
of the theory, which is believed to be given by colorless bound
states of gluons and quarks as, for instance, mesons, baryons and
glueballs. This task is far beyond our capabilities. At intermediate
$Q^2$, what we can state is that this dynamical mass parametrizes
the behaviour of the Greens function of the gluon. As a result of
quantum effects, i.e. the condensation of the mass dimension two
operator, a pole appears in the off-diagonal gluon propagator at the
tree level. Including higher order effects will alter the
propagators behaviour as well as the location of the pole at
physical $Q^2$ (i.e. $Q^2<0$). In the case of the Landau gauge,
higher order calculations showed that the condensate remains stable,
and hence a nonzero mass parameter will remain, see
\cite{Verschelde:2001ia,Browne:2003uv}. This mass parameter will
describe the behaviour of the Greens function at Euclidean $Q^2$.
The presence of a mass parameter does however not necessarily entail
the presence of a pole in the propagator at physical $Q^2$. Using
lattice simulations of the Euclidean propagator, a mass parameter is
found also by fitting at Euclidean $Q^2>0$, but no pole and related
massive particle is implied. Analogously, one should not conclude
from our calculations that the gluon is a massive, physical particle
and that unitarity is violated.

\section*{Acknowledgments.}
The Conselho Nacional de Desenvolvimento Cient\'{\i}fico e
Tecnol\'{o}gico
(CNPq-Brazil), the SR2-UERJ and the Coordena{\c{c}}{\~{a}}o de Aperfei{\c{c}}%
oamento de Pessoal de N{\'\i}vel Superior (CAPES) are gratefully
acknowledged for financial support. D.~Dudal would like to thank the
Theoretical Physics Department of the UERJ for the kind hospitality
where a part of this work was completed, while R.~F.~Sobreiro wants
to acknowledge the warm hospitality at the Department of
Mathematical Physics and Astronomy where another part of this work
was completed.


\begin{thebibliography}{99}
\bibitem{scon}  Y. Nambu, \emph{Phys. Rev. }\textbf{D10} (1974) 4262;\newline
G. 't Hooft, \emph{High Energy Physics EPS Int. Conference,
}Palermo 1975, ed. A. Zichichi;\newline S. Mandelstam, \emph{Phys.
Rept. }\textbf{23} (1976) 245.

\bibitem{'tHooft:1981ht}
G.~'t Hooft, Nucl.\ Phys.\ B {\bf 190} (1981) 455.

\bibitem{Ezawa:bf}
Z.~F.~Ezawa and A.~Iwazaki, Phys.\ Rev.\ D {\bf 25} (1982) 2681.

\bibitem{Suzuki:1989gp}
T.~Suzuki and I.~Yotsuyanagi, Phys.\ Rev.\ D {\bf 42} (1990) 4257.

\bibitem{Hioki:1991ai}
S.~Hioki, S.~Kitahara, S.~Kiura, Y.~Matsubara, O.~Miyamura,
S.~Ohno and T.~Suzuki, Phys.\ Lett.\ B {\bf 272} (1991) 326
[Erratum-ibid.\ B {\bf 281} (1992) 416].

\bibitem{Kronfeld:1987vd}
A.~S.~Kronfeld, G.~Schierholz and U.~J.~Wiese, Nucl.\ Phys.\ B
{\bf 293} (1987) 461.

\bibitem{Kronfeld:1987ri}
A.~S.~Kronfeld, M.~L.~Laursen, G.~Schierholz and U.~J.~Wiese,
Phys.\ Lett.\ B {\bf 198} (1987) 516.

\bibitem{Min:bx}
H.~Min, T.~Lee and P.~Y.~Pac, Phys.\ Rev.\ D {\bf 32} (1985) 440.

\bibitem{Fazio:2001rm}
A.~R.~Fazio, V.~E.~R.~Lemes, M.~S.~Sarandy and S.~P.~Sorella,
Phys.\ Rev.\ D {\bf 64} (2001) 085003.

\bibitem{Amemiya:1998jz}
K.~Amemiya and H.~Suganuma, Phys.\ Rev.\ D {\bf 60} (1999) 114509.

\bibitem{Bornyakov:2003ee}
V.~G.~Bornyakov, M.~N.~Chernodub, F.~V.~Gubarev, S.~M.~Morozov and
M.~I.~Polikarpov, Phys.\ Lett.\ B {\bf 559} (2003) 214.

\bibitem{Schaden:1999ew}
M.~Schaden, hep-th/9909011.

\bibitem{Kondo:2000ey}
K.~I.~Kondo and T.~Shinohara, Phys.\ Lett.\ B {\bf 491} (2000)
263.

\bibitem{Dudal:2002xe}
D.~Dudal and H.~Verschelde, J.\ Phys.\ A {\bf 36} (2003) 8507.

\bibitem{Sawayanagi:2003dc}
H.~Sawayanagi, Phys.\ Rev.\ D {\bf 67} (2003) 045002.

\bibitem{Kondo:2001nq}
K.~I.~Kondo, Phys.\ Lett.\ B {\bf 514} (2001) 335.

\bibitem{Curci:bt}  G.~Curci and R.~Ferrari, Nuovo Cim.\ A \textbf{32}
(1976) 151.

\bibitem{Curci:1976ar}  G.~Curci and R.~Ferrari, Phys.\ Lett.\ B
\textbf{63} (1976) 91.

\bibitem{Dudal:2003pe}
D.~Dudal, H.~Verschelde, V.~E.~R.~Lemes, M.~S.~Sarandy,
R.~F.~Sobreiro, S.~P.~Sorella, M.~Picariello, J.A.~Gracey, Phys.\
Lett.\ B {\bf 569} (2003) 57.

\bibitem{Dudal:2003gu}
D.~Dudal, H.~Verschelde, V.~E.~R.~Lemes, M.~S.~Sarandy,
S.~P.~Sorella and M.~Picariello, Annals Phys.\  {\bf 308} (2003)
62.

\bibitem{Verschelde:2001ia}
H.~Verschelde, K.~Knecht, K.~Van Acoleyen and M.~Vanderkelen,
Phys.\ Lett.\ B {\bf 516} (2001) 307.

\bibitem{Langfeld:2001cz}
K.~Langfeld, H.~Reinhardt and J.~Gattnar, Nucl.\ Phys.\ B {\bf
621} (2002) 131.

\bibitem{Aguilar:2004kt}
A.~C.~Aguilar and A.~A.~Natale, hep-ph/0405024.

\bibitem{Aguilar:2004sw}
A.~C.~Aguilar and A.~A.~Natale, JHEP {\bf 0408} (2004) 057.

\bibitem{Gubarev:2000eu}
F.~V.~Gubarev, L.~Stodolsky and V.~I.~Zakharov, Phys.\ Rev.\
Lett.\  {\bf 86} (2001) 2220.

\bibitem{Gubarev:2000nz}
F.~V.~Gubarev and V.~I.~Zakharov, Phys.\ Lett.\ B {\bf 501}
(2001).

\bibitem{Boucaud:2000nd}
P.~Boucaud, A.~Le Yaouanc, J.~P.~Leroy, J.~Micheli, O.~P\`{e}ne
and J.~Rodriguez-Quintero, Phys.\ Lett.\ B {\bf 493} (2000) 315.

\bibitem{Boucaud:2001st}
P.~Boucaud, A.~Le Yaouanc, J.~P.~Leroy, J.~Micheli, O.~P\`{e}ne
and J.~Rodriguez-Quintero, Phys.\ Rev.\ D {\bf 63} (2001) 114003.

\bibitem{Boucaud:2002nc}
P.~Boucaud, J.~P.~Leroy, A.~Le Yaouanc, J.~Micheli, O.~P\`{e}ne,
F.~De Soto, A.~Donini, H.~Moutarde and J. Rodr{\'\i}guez-Quintero
Phys.\ Rev.\ D {\bf 66} (2002) 034504.

\bibitem{book}  O. Piguet and S.P. Sorella, \textit{Algebraic
Renormalization}, Lect.\ Notes Phys.\  {\bf M28} (1995) 1.

\bibitem{Dudal:2002pq}
D.~Dudal, H.~Verschelde and S.~P.~Sorella, Phys.\ Lett.\ B {\bf
555} (2003) 126.

\bibitem{Dudal:2003np}
D.~Dudal, H.~Verschelde, V.~E.~R.~Lemes, M.~S.~Sarandy,
R.~F.~Sobreiro, S.~P.~Sorella and J.~A.~Gracey, Phys.\ Lett.\ B
{\bf 574} (2003) 325.

\bibitem{Dudal:2003by}
D.~Dudal, H.~Verschelde, J.~A.~Gracey, V.~E.~R.~Lemes,
M.~S.~Sarandy, R.~F.~Sobreiro and S.~P.~Sorella, JHEP {\bf 0401}
(2004) 044.

\bibitem{Parisi:1980jy}
G.~Parisi and R.~Petronzio, Phys.\ Lett.\ B {\bf 94} (1980) 51.

\bibitem{Field:2001iu}
J.~H.~Field, Phys.\ Rev.\ D {\bf 66} (2002) 013013.

\bibitem{Giacosa:2004ug}
F.~Giacosa, T.~Gutsche and A.~Faessler, hep-ph/0408085.

\bibitem{Kondo:1997pc}
K.~I.~Kondo, Phys.\ Rev.\ D {\bf 57} (1998) 7467.

\bibitem{Kondo:2001tm}
K.~I.~Kondo, T.~Murakami, T.~Shinohara and T.~Imai, Phys.\ Rev.\ D
{\bf 65} (2002) 085034.

\bibitem{Barnich:2000zw}
G.~Barnich, F.~Brandt and M.~Henneaux, Phys.\ Rept.\  {\bf 338},
439 (2000).

\bibitem{Verschelde:jj}
H.~Verschelde, Phys.\ Lett.\ B {\bf 351} (1995) 242.

\bibitem{Knecht:2001cc}
K.~Knecht and H.~Verschelde, Phys.\ Rev.\ D {\bf 64} (2001)
085006.

\bibitem{Baulieu:1998kx}
L.~Baulieu and D.~Zwanziger, Nucl.\ Phys.\ B {\bf 548} (1999) 527.

\bibitem{Shinohara:2001cw}
T.~Shinohara, T.~Imai and K.~I.~Kondo, Int.\ J.\ Mod.\ Phys.\ A
{\bf 18} (2003) 5733.

\bibitem{Ellwanger:2002sj}
U.~Ellwanger and N.~Wschebor, Int.\ J.\ Mod.\ Phys.\ A {\bf 18}
(2003) 1595.

\bibitem{Kondo:2003sw}
K.~I.~Kondo, hep-th/0303251.

\bibitem{Browne:2004mk}
R.~E.~Browne and J.~A.~Gracey, Phys.\ Lett.\ B {\bf 597} (2004) 368.

\bibitem{Browne:2003uv}
R.~E.~Browne and J.~A.~Gracey, JHEP {\bf 0311} (2003) 029.

\bibitem{Greensite:1985vq}
J.~Greensite and M.~B.~Halpern, Nucl.\ Phys.\ B {\bf 271} (1986)
379.

\bibitem{Dudal:2003dp}
D.~Dudal, H.~Verschelde, V.~E.~R.~Lemes, M.~S.~Sarandy,
S.~P.~Sorella, M.~Picariello, A.~Vicini and J.~A.~Gracey, JHEP
{\bf 0306} (2003) 003.

\bibitem{Kondo:2000sh}
K.~I.~Kondo, hep-th/0009152.

\bibitem{Deguchi:2003zi}
S.~Deguchi and Y.~Kokubo, Mod.\ Phys.\ Lett.\ A {\bf 18} (2003)
2051.

\bibitem{Freire:2001nd}
F.~Freire, Phys.\ Lett.\ B {\bf 526} (2002) 405.

\bibitem{Cornwall:1981zr}
J.~M.~Cornwall, Phys.\ Rev.\ D {\bf 26} (1982) 1453.


\end{thebibliography}
\end{document}